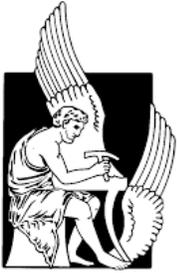

TECHNICAL UNIVERSITY OF CRETE

SCHOOL OF PRODUCTION ENGINEERING AND MANAGEMENT

# Vehicle path and traffic flow optimization via lane changing of automated or semi-automated vehicles on motorways

Georgantas Antonios

Diploma Thesis

Chania 2016

The current Thesis by Georgantas Antonios is approved by the committee:

Papageorgiou Markos Supervisor Professor
Papamichail Ioannis , Assistant Professor
Delis Anargiros , Associate Professor

# Compliments

By completing the present dissertation, I would like to thank Professor Papageorgiou for giving me the opportunity on embarking on the field of Transportation Engineering, as also for the patience and guidance he has shown throughout this period. I also would like to thank the Master student Georgia Perraki, without her help, it wouldn't be possible to finalize this work. Moreover, I would like to thank Dr. Nikolaos Bekiaris-Liberis for his guidance and most importantly I would like to thank Dr. Claudio Roncoli, who has been really very helpful and patient, as also provided me with everything he deemed would be important to grasp relatively to the general meaning of the objective we needed to accomplish. The interaction with Claudio was of paramount importance and set as the foundation for many of the notions I explain in the main body of the Thesis.

# Contents

**Abstract**

Emerging vehicle automation and communication systems (VACS) may contribute to the improvement of vehicles' travel time and the mitigation of motorway traffic congestion on the basis of appropriate control strategies. This work considers the possibility that automated, or semi-automated, vehicles are equipped with devices that perform (or recommend) lane-changing tasks. The lane-changing strategy MOBIL (minimizing overall braking induced by lane changing) has been chosen for its simplicity and ductility, as well as for the reduced number of parameters that need to be specified (namely, politeness factor and threshold). A wide set of simulations, where MOBIL has been implemented within the microscopic traffic simulator Aimsun for a calibrated motorway network (representing a stretch of motorway A12 in the Netherlands), has been performed. Simulations revealed the impact that the choice of different parameters have on the travel time of different vehicles, allowing also to analyse their behaviour with respect to different traffic conditions (without or with traffic congestion).

## Περίληψη

Η εντεινόμενη εξέλιξη στο χώρο του αυτοματισμού και της διασύνδεσης των οχημάτων μέσω συστημάτων VACS (Vehicle Automation and Communication Systems), μπορεί να συντελέσει θετικά στην ελαχιστοποίηση της κυκλοφοριακής συμφόρησης, βάση κατάλληλων στρατηγικών κυκλοφοριακού ελέγχου. Η παρούσα εργασία εξετάζει τη δυνατότητα αυτή με ευφυή συστήματα που προτείνουν σε αυτόματα ή ημιαυτόματα οχήματα την αλλαγή λωρίδας σε δίκτυα αυτοκινητοδρόμων. Για την υλοποίηση της εργασίας γίνεται χρήση του μοντέλου MOBIL (minimizing overall braking induced by lane change) που είναι ένα γενικό αλλά απλό μοντέλο αλλαγής λωρίδας, το οποίο εκτιμά την χρησιμότητα αλλαγής λωρίδας για την πορεία ενός οχήματος, αλλά και τις επιβραδύνσεις που μπορεί αυτή να επιφέρει σε άλλα οχήματα και αποφασίζει, ανάλογα με τις τιμές που θα δοθούν σε δύο παραμέτρους, για την καταλληλότητα αλλαγής λωρίδας ή όχι. Η λογική αυτή υλοποιήθηκε στο λογισμικό μικροσκοπικής προσομοίωσης ( Aimsun ) και εφαρμόστηκε, σε δίκτυο αυτοκινητοδρόμου στην Ολλανδία (που περιλαμβάνει δυο ράμπες εισόδου, δυο ράμπες εξόδου και ενα lane-drop) σε συνδυασμό με το car following model IDM (Intelligent Driver Model). Πραγματοποιήθηκε βελτιστοποίηση των παραμέτρων του MOBIL, όπως ο συντελεστής ευγενίας και το απαιτούμενο κατώφλι ώστε να πραγματοποιηθεί η αλλαγή λωρίδας με στόχο, αφενώς την ελαχιστοποίηση του χρόνου παραμονής των αυτοκινήτων (που είναι εναρμονισμένα με την λογική του MOBIL) και αφετέρου την ελαχιστοποίηση του χρόνου παραμονής όλων των υπόλοιπων οχημάτων του δικτύου.

# Chapter 1

# Introduction

The lane-changing model is an important component of microscopic traffic simulation tools. With the increasing popularity of these tools, a number of lane-changing models have been proposed and implemented in various simulators in recent years [2]. Collective phenomena such as traffic instabilities and the spatio-temporal dynamics of congested traffic can be well understood within the scope of single-lane traffic models. In order to produce a more realistic description of the lane-changing impact in more complicated situations, it is necessary to examine its behaviour within a multi-lane modeling framework, allowing faster vehicles to improve their driving condition by passing slower vehicles. The modeling of lane changes is typically considered a multistep process. On a strategic level, the driver knows about his or her route in a network, which influences the lane choice, for example, with regard to lane blockages, on-ramps, off-ramps, or other mandatory merges . In the tactical stage, an intended lane change is prepared and initiated by advance acceleration or deceleration by the driver and possibly by cooperation of drivers in the target lane . Finally, in the operational stage, one determines if an immediate lane change is both safe and desirable. For the sake of safety, we have emulated a gap-acceptance model, which takes into account the leader and the follower of our current vehicle in the target lane.In general, there are many other inclinations towards lane changing optimization. Our main motivation in this Thesis, is to evaluate the efficiency of the attached network, by incorporating automated or semi-automated vehicles which are adjusted with a controller named MOBIL. We then will detect how automated vehicles affect the network and the other individual vehicles of the network and ameliorate the travel time and traffic state not only of automated vehicles, but also of vinicity vehicles within our calibrated network. As far as the lane changing process is concerned, we will be able to test the automated behaviour of vehicles by employing a model that describes the rational decision to change lanes and therefore deals only with the operational decision process. When a lane change is considered, it is assumed that a driver makes a trade-off between the expected own advantage and the disadvantage imposed on other drivers. In particular, the current model includes the follower in the target lane in the decision process. For a driver considering a lane change, the subjective utility of a change increases with the gap to the new leader in the target lane. However, if the velocity of this leader is lower, it may be favorable to stay in the present lane despite the smaller gap. A criterion for the utility including both situations is the difference in the accelerations after and before the lane-change. In this work, therefore, it is proposed that the utility function be consideration of the difference in vehicle accelerations (or decelerations) after a lane change, calculated with an underlying microscopic longitudinal traffic model. The higher the acceleration in a given lane,the nearer it is to the ideal acceleration on an empty road and the more attractive it is to the driver. Therefore, the basic idea of the proposed lane-changing model is to formulate the anticipated advantages and disadvantages of a prospective lane change in terms of single-lane accelerations.We need to take into account, the advantage or disadvantage of the followers via a "politeness parameter. By adjusting this parameter, the motivations for lane changing can be varied from purely egoistic to more altruistic behavior. In particular, there exists a value at which lane changes are carried out only if they increase the combined accelerations of the lane-changing driver and all affected neighbors. This strategy can be paraphrased by the phrase "minimizing overall braking induced by lane changes "(MOBIL). [3]

The remaining part of the Thesis is dedicated in explaining the logic behind MOBIL model and the realization of MOBIL strategy within our simulated network. I would describe the microscopic car following model IDM (Intelligent Driver Model) as an accompanying model for MOBIL controller.

As a point of reference I use a calibrated motorway network in a microscopic environment.[10]. Afterwards, we want to check the travel time estimated for different array of vehicles injected with MOBIL controller.

# Chapter 2

# Microscopic Modeling of Traffic Flows

Microscopic modeling of traffic flows is based on the description of the motion of each individual vehicle composing the traffic stream. This implies modeling the actions – e.g., acceleration, decelerations, and lane changes – of each driver in response to the surrounding traffic. According to May, theories describing how one vehicle follows another vehicle were developed primarily in the 1950s and 1960s, after the pioneering development of car-following theories by Reuschel(1950) and Pipes(1953) . Pipes' work, based on the concept of distance headway, characterizes the motion of vehicles in the traffic stream as following rules suggested in the California Motor Vehicle Code, namely "A good rule for following another vehicle at a safe distance is to allow yourself at least the length of a car between your vehicle and the vehicle ahead for every ten miles per hour of speed at which you are traveling.[4]

## 2.1 Advancements

Much more extensive research was undertaken in the late 1950s by the General Motors Group based on comprehensive field experiments and the development of the mathematical theory bridging micro and macro theories of traffic flows. This research led to the formulation of the car-following models as a form of stimulus-response equation (Gerlough and Huber, 1975), where the response is the reaction of a driver to the motion of the vehicle immediately preceding him in the traffic stream. The response is always to accelerate or decelerate in proportion to the magnitude of the stimulus at time t and begins after a time lag T, the reaction time of the follower. [4]

## 2.2 Lane-changing in motorways

One of the major problems faced when proposing alterations to a road is to estimate in advance the likely effect of the alterations, and to determine possible side effects. If the traffic is so heavy that lane-changing is not possible or can only be performed by slow mergings at over- saturated bottlenecks, analytical methods of estimation can be devised. With lighter traffic the ability to change lanes, and the interaction between various features of the driving environment, make analytical techniques impossible in many urban situations. Factors such as turning movements, public transport and lane closures interact with each other and the general traffic, and the results of change are hard to predict. In these situations computer simulation represents a particularly straightforward approach to the problem.At the level of the individual driver the effects of the various features of the road and traffic can produce reactions in two dimensions. The driver can accelerate or brake in an effort to maintain his desired speed without running into the vehicle ahead, or can change lanes.Car-following models are relatively unaffected by what constitutes the previous vehicle; a fixed obstruction or a stop line at an intersection where the lights are red or amber constitute acceptable substitutes for a real vehicle. Nor do changes in desired speed, acceleration and braking with curves and gradients affect the models.Lane changing, however, is more complex, because the decision to change lanes depends on a number of objectives, and at times these may conflict. For instance, a driver may be in the rightmost lane and wish to turn right within 50 metres but still have to change lanes to the left to avoid a breakdown. Or, a driver may be able to increase his speed in the short term by changing lanes, but in doing so become trapped behind a slow heavy vehicle and lose more than any temporary advantage he might gain. Thus, in coming to a decision concerning lane changing, a driver must be able to reconcile his short-term and long-term aims.[5]

## 2.3 Lane-changing model MOBIL

When we refer to MOBIL, we imply the phrase "minimizing overall braking induced by lane changes ". This model considers accelerations as utility functions. Each utility is attached to the adjacent lanes of the lane that our vehicle is currently situated.The higher the acceleration in a given lane,the nearer it is to the ideal acceleration on an empty road and the more attractive it is to the driver. Therefore, the basic idea of the proposed lane-changing model is to formulate the anticipated advantages and disadvantages of a prospective lane change in terms of single-lane accelerations.

The formulation in terms of accelerations of a longitudinal model has several advantages. First, assessment of the traffic situation is transferred to the acceleration function of the car-following model, which allows for a compact and general model formulation with only a small number of additional parameters. In contrast to the classical gap-acceptance approach, critical gaps are not taken into account explicitly. Second,it is ensured that both longitudinal and lane-changing models are consistent with each other. For example, if the longitudinal model is collision-free, the combined models will be accident-free as well.Third, any complexity of the longitudinal model such as anticipation is transferred automatically to a similarly complex lane-changing model. Finally, the braking deceleration imposed on the new follower in the target lane to avoid accidents is an obvious measure for safety. Thus, safety and motivational criteria can be formulated in a unified way.[3]

A factor that requires close attention, is the politeness factor $p$. By adjusting this parameter, the motivations for lane changing can be varied from purely egoistic to more altruistic behaviour. In the following, the concept discussed here is referred to with this acronym regardless of the value of the politeness parameter. By the politeness factor, two common lane-changing patterns can be modeled. First, most drivers do not change lanes for a marginal advantage if this change obstructs other drivers in addition to a common safety condition. Second, in countries with asymmetric lane-changing rules, aggressive drivers may induce the lane change of a slower driver in front of them to the faster lane, which is dedicated to passing, so that the slower lane will no longer be obstructed.

In this Thesis, we are going to focus only in symmetric lane changing strategies.

In reality there is a distinction between manual vehicles, which are being driven by a human, as also automated or semi-automated vehicles, which merely or completely allow the advisory system of the vehicle to perform lane changing in the most efficient way possible. However, this isn't exactly the case in the simulated world, where several aspects need to be taken into account. More specifically, there are mathematical expressions and equations, which try to emulate what reality is within a constrained environment, so as to respond to reality in the most suitable way. For this Thesis, we have incorporated IDM car following model and MOBIL lane changing model, both of them in a unified essence can perform as actuators of the automated lane changing in motorways. MOBIL can be considered as a controller (with the broader meaning), that takes some measurements and decisions according to the prospective accelerations of the putative leader and follower in the target lane.

Due to its simplicity, MOBIL is an ideal tool for performing automated lane changing. As we will see later on, MOBIL takes into account the longitudinal accelerations in the original lane, as also the accelerations computed based on the putative leader and follower in the target lane.

Most time-continuous microscopic single-lane traffic models describe the motion of single driver–vehicle units *a* as a function of their own velocity $v_\alpha$, the bumper-to-bumper distance $s_\alpha$ to the front vehicle $(\alpha - 1)$ and the relative velocity $\Delta v_\alpha = v_\alpha - v_{\alpha-1}$. The acceleration of these car-following models is of the following general form:

$$a_\alpha = \frac{dv_\alpha}{dt} = a(s_\alpha, v_\alpha, \Delta v_\alpha) \tag{2.1}$$

A generalization to models taking into account more than one predecessor or to models with explicit reaction time is straightforward.

Based on equation (2.1), we can deduce, that this form can be applied not only in manual vehicles, but also in automated vehicles as well (as those referred previously). This is the case, since the computation of accelerations to retrieve the behaviour of a vehicle, is common for several controllers proposed for automated vehicles such as ACC ( Adaptive Cruise Control). Some other controllers along with ACC can be found with greater detail at: [11],[12], [13], [15]. This is an advantage of MOBIL, because any automated vehicle can be attached with the MOBIL controller (because of the Incentive Criterion inequality (2.2)). This lane changing model as will mentioned later on, computes only longitudinal accelerations and it's entire logic is structured based on (2.1). There are also other car following models apart from IDM, which emulate the ACC ([14]).

A specific lane change (e.g., from the center lane to the median lane as shown in Figure 2.1 depends generally on the two following vehicles in the current and the target lanes, respectively. To formulate the lane-changing criteria, the following notation is used: for a vehicle c considering a lane change, the successive vehicles in the target and current lanes are represented by n and o, respectively. The acceleration ac denotes the acceleration of vehicle c on the actual lane, and $\tilde{\alpha}_c$ refers to the situation in the target lane, that is, to the new acceleration of vehicle c in the target lane. Likewise,$\tilde{\alpha}_0$ and $\tilde{\alpha}_n$ denote the acceleration of the old and new followers after the lane change of vehicle c.

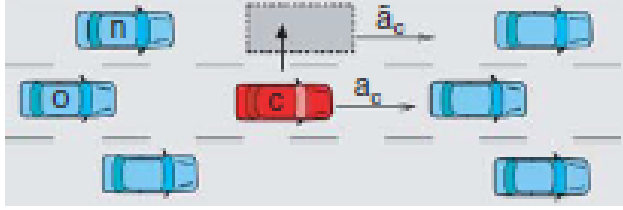

Figure 2.1: Nearest neighbors of central vehicle c considering lane change to the left (new and old successors are denoted n and o respectively:accelerations after possible change are denoted with a tilde).

### 2.3.1 Incentive Criterion for Symmetric Lane-Changing Rules

The incentive criterion typically determines if a lane change improves the individual local traffic situation of a driver. In the current model, the incentive criterion is generalized to include the immediately affected neighbors as well. The politeness factor p determines to which degree these vehicles influence the lane-changing decision. This logic can be depicted in the following inequality:

$$\tilde{\alpha}_c - \alpha_c + p(\tilde{\alpha}_n - \alpha_n + \tilde{\alpha}_o - \alpha_o) > \Delta\alpha_{th} \tag{2.2}$$

The first two terms denote the advantage (utility) of a possible lane change for the driver where $\tilde{\alpha}_c$ refers to the new acceleration for vehicle c after a prospective lane change. The considered lane change is favorable if the driver can accelerate more, that is, go faster in the new lane. The third term with the politeness factor p is the main innovation in this model. It denotes the total advantage (acceleration gain or loss, if negative) of the two immediately affected neighbors, weighted with p. Finally, the switching threshold $\Delta\alpha_{th}$ on the right-hand side of (2.2) models a certain inertia and prevents lane changes if the overall advantage is only marginal compared with a "keep lane" directive. In summary, the incentive criterion is fulfilled if the own advantage (acceleration gain) is higher than the weighted sum of the disadvantages (acceleration losses) of the new and old successors and the threshold $\Delta\alpha_{th}$. In fact, the incentive criterion in (2.2) automatically includes a safety component for the lane changing vehicle. Even for the most aggressive parameter settings ($p = 0$ and $\Delta\alpha_{th} = 0$) lanes are only changed if, in the new lane, the acceleration is higher or, equivalently, the necessary braking deceleration is lower than in the current lane. Consequently, Criterion (2.2) can only be true if the new lane is safer than the old lane. The only requirement for the acceleration model is that, in dangerous situations, it should return a braking deceleration that increases as the situation becomes more critical, a condition that any reasonable acceleration model should fulfill.] It should be noted that the threshold $\Delta \alpha_{th}$ influences the lane-changing behavior globally, whereas the politeness parameter affects the local lane-changing behavior depending on the involved neighbors.[3]

The generalization to traffic in more than two lanes per direction is straightforward. If, for a vehicle in a center lane, the incentive criterion is satisfied for both neighboring lanes, the change is performed to the lane in which the incentive is larger.

Since the disadvantages of other drivers and the own advantage are balanced via the politeness factor p, the lane-changing model contains typical strategic features of classical game theory. The value of p can be interpreted as the degree of altruism. It can vary from $p = 0$ (for selfish lane-hoppers) to $p > 1$ for altruistic drivers who do not change if that would cause the overall traffic situation to deteriorate considering followers, whereas they would perform even disadvantageous lane changes if that change improved the situation of the followers sufficiently.

One specific key-point in the whole MOBIL infrastructure is , that lane changes are only performed when they increase the sum of accelerations of all involved vehicles, which correspond to the concept of minimizing overall braking induced by lane changes (MOBIL) in the ideal sense.

We can incorporate the MOBIL logic in the following Figure 2.2 , combining different thresholds and politeness factors for more aggressive and more altruistic driving behaviour.

## 2.4 Car following models

Car following models feauture traffic dynamics from the perspective of individual driver-vehicle units(expresses the fact that the driving behavior depends not only on the driver but also on the acceleration and braking capabilities of the vehicle).[6]

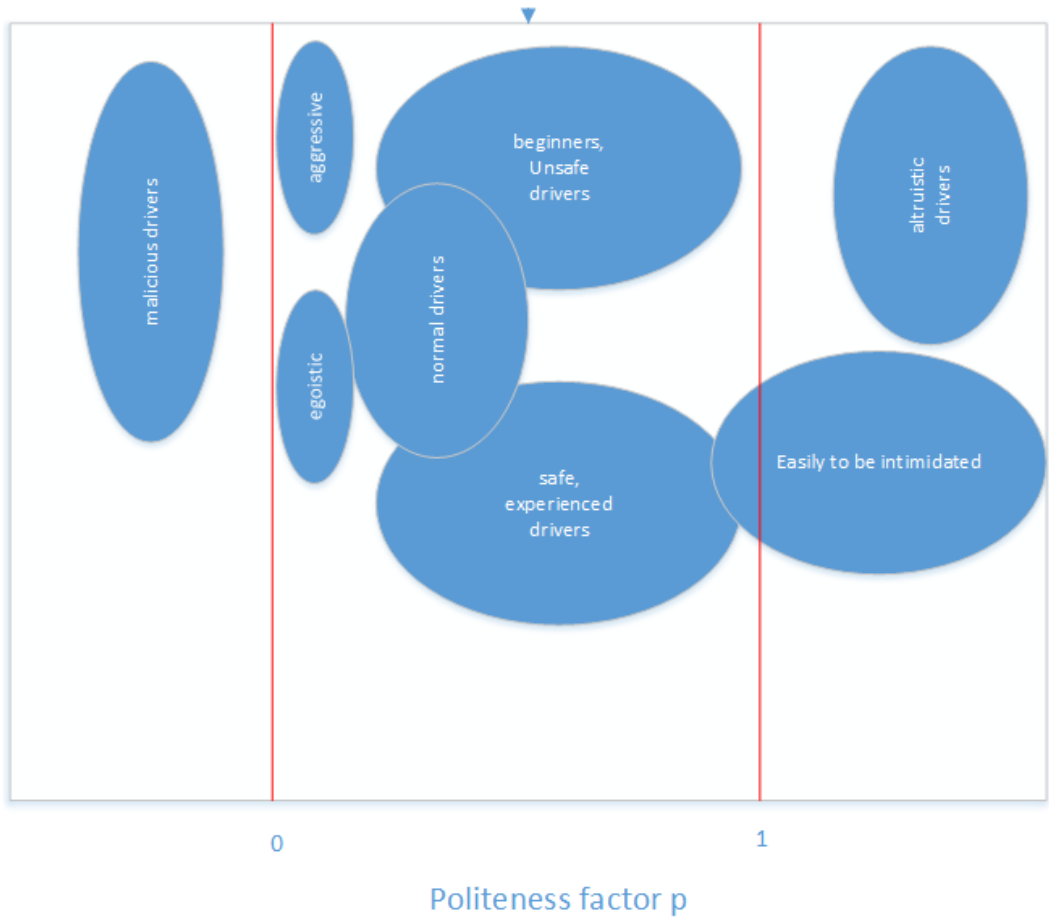

Figure 2.2: Classification of different driver types with respect to the politeness factor. While the safety criterion prevents critical lane changes and collisions, the incentive criterion also takes into account the (dis-) advantages of other drivers associated with a lane change. Most other lane-changing models implicitly adopt an egoistic behavior (p = 0), . For p = 1, lane changes always lead to an increase of the average accelerations of the vehicles involved (MOBIL principle).

In a strict sense, car-following models describe the driver's behavior only in the presence of interactions with other vehicles, while free traffic flow is described by a separate model. In a more general sense, car-following models include all traffic situations such as car-following situations, free traffic, and also stationary traffic. In this case we say that the microscopic models is complete:

A car-following model is complete if it is able to describe all situations including acceleration and cruising in free traffic, following other vehicles in stationary and non-stationary situations, and approaching slow or standing vehicles, and red traffic lights.

The first car-following models were proposed more than fifty years ago by Reuschel (1950), and Pipes (1953). These two models already contained one essential element of modern microscopic modeling: The minimum bumper-to-bumper distance to the leading vehicle (also known as the "safety distance") should be proportional to the speed. This can be expressed equivalently by requiring that the time gap should not be below a fixed safe time gap.[6]

Each driver-vehicle combination $\alpha$ is described by the state variables location $x_\alpha(t)$ (position of the front bumper along the arc length of the road, increasing in driving direction), and speed $\dot{v}_\alpha(t)$ as a function of the time $t$, and by the attribute "vehicle length" $l_\alpha$. Depending on the model, additional state variables are required, for example, the acceleration
$\dot{v}_\alpha = dv/dt$, or binary activation-state variables for brake lights or indicators. We define the vehicle index $\alpha$ such that vehicles pass a stationary observer (or detector) in ascending order, i.e., the first vehicle has the lowest index (cf. Figure 2.3). Notice that this implies that the vehicles are numbered in descending order with respect to their location $x$. From the vehicle locations and lengths, we obtain the (bumper-to-bumper) distance gaps, which(together with the vehicle speeds) constitute the main input of the microscopic models.

$$s_\alpha = x_{\alpha-1} - l_{\alpha-1} - x_\alpha = x_l - l_l - x_\alpha \tag{2.3}$$

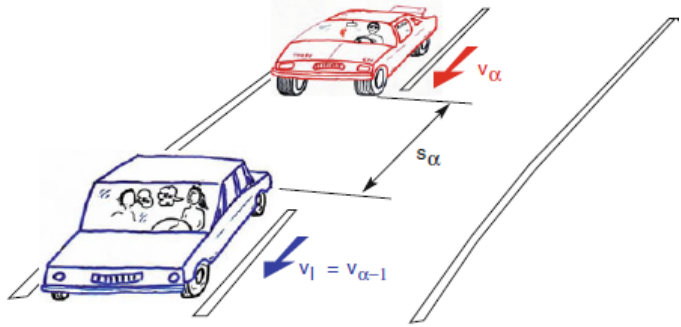

Figure 2.3: Definition of state variables of car-following models

The minimal models (and many of the more realistic models of ) describe the response of the driver as a function of the gap $s_\alpha$ to the lead vehicle, the driver's speed $v_\alpha$, and the speed $v_l$ of the leader. *In continuous-time models*, as the ones we are going to examine later on in the procedure of lane changing, the driver's response is directly given in terms of an acceleration function $\alpha_{mic}(s, v, v_l)$ leading to a set of *coupled ordinary differential equations* of the form

$$\dot{x}_\alpha(t) = \frac{dx_\alpha(t)}{dt} = v_\alpha(t), \tag{2.4}$$

$$\dot{v}_\alpha(t) = \frac{dv_\alpha(t)}{dt} = \alpha_{mic}(s_\alpha, v_\alpha, v_l) = \tilde{\alpha}_{mic}(s_\alpha, v_\alpha, \Delta v_\alpha). \tag{2.5}$$

In most acceleration functions, the speed $v_l$ of the leader enters only in form of the speed difference (approaching rate)

$$\Delta v_\alpha = v_\alpha - v_{\alpha-1} = v_\alpha - v_l \tag{2.6}$$

The corresponding models can be formulated more concisely in terms of the alternative acceleration function

$$\tilde{\alpha}_{mic}(s, v, \Delta v) = \alpha_{mic}(s, v, -\Delta v) \qquad (2.7)$$

Taking the time derivative of (2.3), we can reformulate (2.4) by

$$\dot{s}_\alpha(t) = \frac{ds_\alpha(t)}{dt} = v_l(t) - v_\alpha(t) = -\Delta v_\alpha(t), \qquad (2.8)$$

The set of Eqs. (2.5) and (2.8) can be considered as the generic formulation of most time-continuous car-following models. In this formulation, the coupling between the gap $s_\alpha$ and the speed $v_\alpha$ as well as the coupling between the speed $v_\alpha$ and the speed $v_l$ of the leader becomes explicit.

## 2.5 Car following models based on Driving Strategies

In contrast with the outline of generic car following models, these models incorporate realistic driving conditions as well in the process of lane changing.In particular, some of the notions that should always be taken into account are : keeping a "safe" distance from the leading vehicle, driving at a desired speed , or preferring accelerations to be within a comfortable range. Additionally, kinematic aspects are taken into account, such as the quadratic relation between braking distance and speed.In order to deploy the MOBIL model with an underlying car following model, we formulate the Intelligent Driver Model (IDM). IDM uses the same input variables as the sensors of adaptive cruise control (ACC) systems, and produces a similar driving behaviour.[6]

### 2.5.1 Model Criteria

The models introduced in this chapter are formally identical to the generic car following models presented above. They are defined by an acceleration function $\alpha_{mic}$ (see Eq.(2.5) or a speed function $v_{mic}$. In contrast to the minimal models, the acceleration or speed functions encoding the driving behavior should at least model the following aspects:

1. The acceleration is a strictly decreasing function of the speed. Moreover, the vehicle accelerates towards a desired speed $v_0$ if not constrained by other vehicles or obstacles:

$$\frac{\partial \alpha_{mic}(s,v,v_l)}{\partial v} < 0, \quad \lim_{s \to \infty} \alpha_{mic}(s,v_0,v_l) = 0 \quad \text{for all} \quad v_l \tag{2.9}$$

2. The acceleration is an increasing function of the distances to the leading vehicle:

$$\frac{\partial \alpha_{mic}(s,v,v_l)}{\partial s} \geq 0 \quad \lim_{s \to \infty} \frac{\partial \alpha_{mic}(s,v,v_l)}{\partial s} = 0 \quad \text{for all} \quad v_l \tag{2.10}$$

The inequality becomes an equality if other vehicles or obstacles (including "virtual" obstacles such as the stopping line at a red traffic light) are outside the interaction range and therefore do not influence the driving behavior. This defines the *free-flow acceleration*

$$\alpha_{free}(v) = \lim_{s \to \infty} \alpha_{mic}(s,v,v_l) =\geq \quad \alpha_{mic}(s,v,v_l) \tag{2.11}$$

3. The acceleration is an increasing function of the speed of the leading vehicle. Together with requirement (1), this also means that the acceleration decreases (the deceleration increases) with the speed of approach to the lead vehicle (or obstacle):

$$\frac{\partial \tilde{\alpha}_{mic}(s,v,\Delta v)}{\partial \Delta v} \leq 0 \quad or \quad \frac{\partial \alpha_{mic}(s,v,v_l)}{\partial v_l} \geq 0, \quad \lim_{s \to \infty} \frac{\partial \alpha_{mic}(s,v,v_l)}{\partial v_l} = 0 \quad (2.12)$$

Again, the equality holds if other vehicles (or obstacles) are outside the interaction range.

4. A minimum gap *(bumper-to-bumper distance)* $s_0$ to the leading vehicle is maintained (also during a standstill). However, there is no backwards movement if the gap has become smaller than $s_0$ by past events:

$$\alpha_{mic}(s,0,v_l) = 0 \quad \text{for all} \quad v_l \geq 0, \quad s_0 \leq 0 \quad (2.13)$$

A car-following model meeting these requirements is complete in the sense that it can consistently describe all situations that may arise in single-lane traffic. Particularly, it follows that (i) all vehicle interactions are of finite reach, (ii) following vehicles are not "dragged along",

$$\alpha_{mic}(s,v,v_l') \leq \alpha_{mic}(inf,v,v_l) = \alpha_{free} \quad \text{for all} \quad s,v,v_l,v_l' \quad (2.14)$$

This means that the model possesses a unique steady-stateow-density relation, i.e., a fundamental diagram.1 These conditions are necessary but not sufficient. For example, when in the car following regime (steady-state congested traffic), the time gap to the leader has to remain within reasonable bounds (say, between 0.5 and 3s). Furthermore, the acceleration has to be constrained to a "comfortable "range (e.g.,$2m/s^2$), or at least, to physically possible values.
Particularly,when approaching the leading vehicle,the quadratic relation between braking distance and speed has to be taken into account. Finally,any car following model should allow instabilities and thus the emergence of "stop-and-go" traffic waves,but should not produce accidents,i.e.,negative bumper to-bumper gaps $s<0$.

### 2.5.2 Car following model IDM

The time-continuous Intelligent Driver Model (IDM) is probably the simplest complete and accident-free model producing realistic acceleration profiles and a plausible behaviour in essentially all single-lane traffic situations.

The IDM is derived from a list of basic assumptions (first-principles model).Thus certain requirements must be met:

1. The acceleration fulfills the general conditions (2.9)–(2.13) for a complete model.

2. The equilibrium bumper-to-bumper distance to the leading vehicle is not less than a "safe distance" $s_0 + vT$ where $s_0$ is a minimum (bumper-to-bumper) gap, and T the (bumper-to-bumper) time gap to the leading vehicle.

3. An intelligent braking strategy controls how slower vehicles (or obstacles or red traffic lights) are approached:

    - Under normal conditions, the braking maneuver is "soft", i.e., the deceleration increases gradually to a comfortable value $b$, and decreases smoothly to zero just before arriving at a steady-state car-following situation or coming to a complete stop.
    - In a critical situation, the deceleration exceeds the comfortable value until the danger is averted. The remaining braking maneuver (if applicable) will be continued with the regular comfortable deceleration $b$.

4. Transitions between different driving modes (e.g., from the acceleration to the car-following mode) are smooth. In other words, the time derivative of the acceleration function, i.e., the jerk $J$, is finite at all times. This is equivalent to postulating that the acceleration function $\alpha_{mic}(s, v, v_l)$ (or $\tilde{a}_{mic}(s, v, \Delta v)$) is continuously differentiable in all three variables.

5. The model should be as parsimonious as possible. Each model parameter should describe only one aspect of the driving behavior (which is favorable for model calibration). Furthermore, the parameters should correspond to an intuitive interpretation and assume plausible values.

#### 2.5.2.1 Mathematical Description

The required properties are realized by the following acceleration equation:

$$\dot{v} = \alpha \left[ 1 - \left( \frac{v}{v_0} \right)^\delta - \left( \frac{s*(v, \Delta v)}{s} \right)^2 \right] \quad (2.15)$$

The acceleration of the Intelligent Driver Model is given in the form $\tilde{\alpha}_{mic}(s, v, \Delta v)$ and consists of two parts, one comparing the current speed $v$ to the desired speed $v_0$, and one comparing the current distance s to the desired distance $s$. The desired distance

$$s*(v, \Delta v) = s_o + max\left(0, vT + \frac{v \Delta v}{2\sqrt{(ab)}}\right). \quad (2.16)$$

has an equilibrium term $s_0 + vT$ and a dynamical term $v \Delta v / 2 \sqrt{(ab)}$

### 2.5.3 Imposed Parameters

We can easily interpret the model parameters by considering the following three standard situations:

- When *accelerating on a free road from a standstill*, the vehicle starts with the maximum acceleration $\alpha$. The acceleration decreases with increasing speed and goes to zero as the speed approaches the desired speed $v_0$. The exponent $\delta$ controls this reduction: The greater its value, the later the reduction of the acceleration when approaching the desired speed.

- When *following a leading vehicle*, the distance gap is approximately given by the safety distance $s_0 + vT$. The safety distance is determined by the time gap T plus the minimum distance gap $s_0$.

- When *approaching slower or stopped vehicles*, the deceleration usually does not exceed the comfortable deceleration b. The acceleration function is smooth during transitions between these situations.

Since the IDM has no explicit reaction time and its driving behavior is given in term of a continuously differentiable acceleration function, the IDM describes more closely the characteristics of semi-automated driving by adaptive cruise control (ACC) than that of a human driver. However, it can easily be extended to capture human aspects like estimation errors, reaction times, or looking several vehicles ahead.

In contrast to other car following models, the IDM explicitly distinguishes between the safe time gap $T$, the speed adaptation time $\tau = v_0/\alpha$, and the reaction time $T_r$. This allows us not only to reflect the conceptual difference between ACCs and human drivers in the model, but also to differentiate between more nuanced driving styles such as "sluggish, yet tailgating" (high value of $\tau = v_0/(\alpha$, low value for T) or "agile, yet safe driving" (low value of $\tau = v_0/(\alpha$, normal value for T, low value for b).

## Chapter 3

# Application of MOBIL to a simulation network

The potential effectiveness of the underlying model MOBIL is investigated in a modified version of a road section of the freeway *A*20 from Rotterdam to Gouda in the Netherlands (the original network is within [8]) . This road section was selected for its infrastructure (ramps, lane drop, relatively closely spaced detectors) and for the congestion patterns that occur (standing queue at the lane drop).
In order to accomodate for MOBIL, we attach to it the car following model we described above (IDM).

## 3.1 Network description

The current network we are going to emulate [10] is about 9 km in length, comprises 2 on-ramps and 2 off-ramps. The stretch includes one lane-drop located at 3.6km (Figure 3.3) .There are 24 detectors placed rather homogeneously at a distance of 300m on average, as illustrated in Figure 3.3. Furthermore, there is a flow detector at off -ramp "Moordrecht".
Measurements from the morning rush hour of Wednseday, May 26, 2010 are incrorporated, where there is a strong congestion initiated at 6:30 AM. This behaviour coincides with the fact, of the increased flow entering from on-ramp "Nieuwerkerk a/d IJssel". The congestion dissapears after 8:10 due to the reduced demand.
The network is space-discretised with N = 23 segments, where each segment is delimited by a pair of detectors. Based on this space-discretisation , on ramps are placed within segments 11,18 whereas off-ramps are placed within segments 8,15.

Figure 3.1: A graphical representation of the considered stretch of highway A20 from Rotterdam to Gouda, the Netherlands [10].

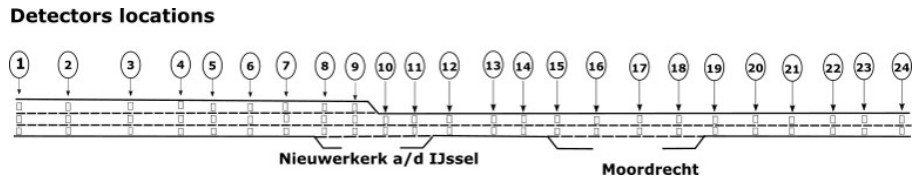

## 3.2 Aimsun Simulator

Microscopic traffic simulators are able to reproduce to a significant level of accuracy the observed traffic conditions in a broad variety of circumstances makes them crucial for the analysis of traffic congestion patterns [9] . For the sake of the emulation of the network we are going to implement, we incorporate the Aimsun software (Advanced Interactive Microscopic Simulator for Urban and Non-Urban Networks).
Microscopic traffic simulators are simulation tools that realistically emulate the flow of individual vehicles through a road network. Aimsun is based on the family of car-following, lane changing and gap acceptance models to model the vehicle's behavior. Aimsun is a proven tool for aiding transportation feasibility studies. This is not only due to its ability to capture the full dynamics of time dependent traffic phenomena, but also because they are capable of using behavioral models that can account for drivers' reactions, when exposed to Intelligent Transport Systems (ITS).The way in which the the car-following model has been implemented in Aimsun takes into account the additional constraints on the breaking capabilities of the vehicles. The implementation tries also to capture the empirical evidence that driver behaviour depends also on local circumstances (i.e. acceptance of speed limits on road sections, influence of grades, friction with drivers in adjacent lanes, and so on). This is done in Aimsun by means of model parameters whose values, calculated at each simulation step, depend on the current circumstances and conditions at each part of the road network.
One feature that is necessary for the representation of the network is GETRAM

(Generic Environment for Traffic Analysis and Modeling), which is responsible for :

- The ability to accurately represent any road network geometry: An easy to use Graphic User Interface, that can use existing digital maps of the road network allows the user to model any type of traffic facility.

- Detailed modeling of the behavior of individual vehicles. This is achieved by employing sophisticated and proven car following and lane changing models that take into account both global and local phenomena that can influence each vehicle's behavior.

- Animated 2D and 3D output of the simulation runs. This is not only a highly desirable feature but can also aid the analysis and understanding of the operation of the system being studied and can be a powerful way to gain widespread acceptance of complex strategies.

### 3.2.1 Aimsun Scenario

A scenario is a microscopic simulation model of a traffic network, or a sub-network of a large network, in which a traffic problem has been identified: the so-called problem network. The model input reproduces to a great degree of accuracy the traffic demand in the problem network for the time period for which the traffic problem has been identified, as well as the current operational conditions in the road network (i.e. current traffic control at signalized intersections, reductions of capacity at specific parts of the network by road works, incidents, and so on). The analysis of the scenario consists on a set of simulation experiments whose purpose is to help the traffic manager to develop and evaluate the impacts of the single actions or combination of actions, consisting of situation related measures (i.e. re-routings and/or speed control using Variable message Panels (VMS), changes in control, an so on), with the objective of alleviating or eliminating the traffic problem identified. This concept of action composed by the various situation-related measures is called a strategy [9].

## 3.3 Traffic Scenario

The previously mentioned network (Figure 3.3) was calibrated with real data (obtained from detectors of a stretch of the highway A20 from Rotterdam to Gouda in the Netherlands) through Aimsun software [10] , fraction of the network referred to [11] . As we discussed in Chapter 3, simulated world tries to approach realistic groundtruths. This means, that for manual vehicles some models to provide the background equations need to be established. These comprise IDM car following model and Gipps generic lane changing model [16]. In order to provide a thorough explanation of our simulated network, since it incorporated lane-drops, some rules need to be developed for the lane-drop area of the network. Automated vehicles depend on the IDM and MOBIL formulation as described in Chapter 3. As These data were taken at morning rush hour of Wednesday, May 26, 2010 . In order to have a more concrete understanding of the calibrated results, we attach the following figure of some 3D plots(one for each lane of the network) illustrating the traffic conditions from 5 AM to 9 AM (Figure 3.2):

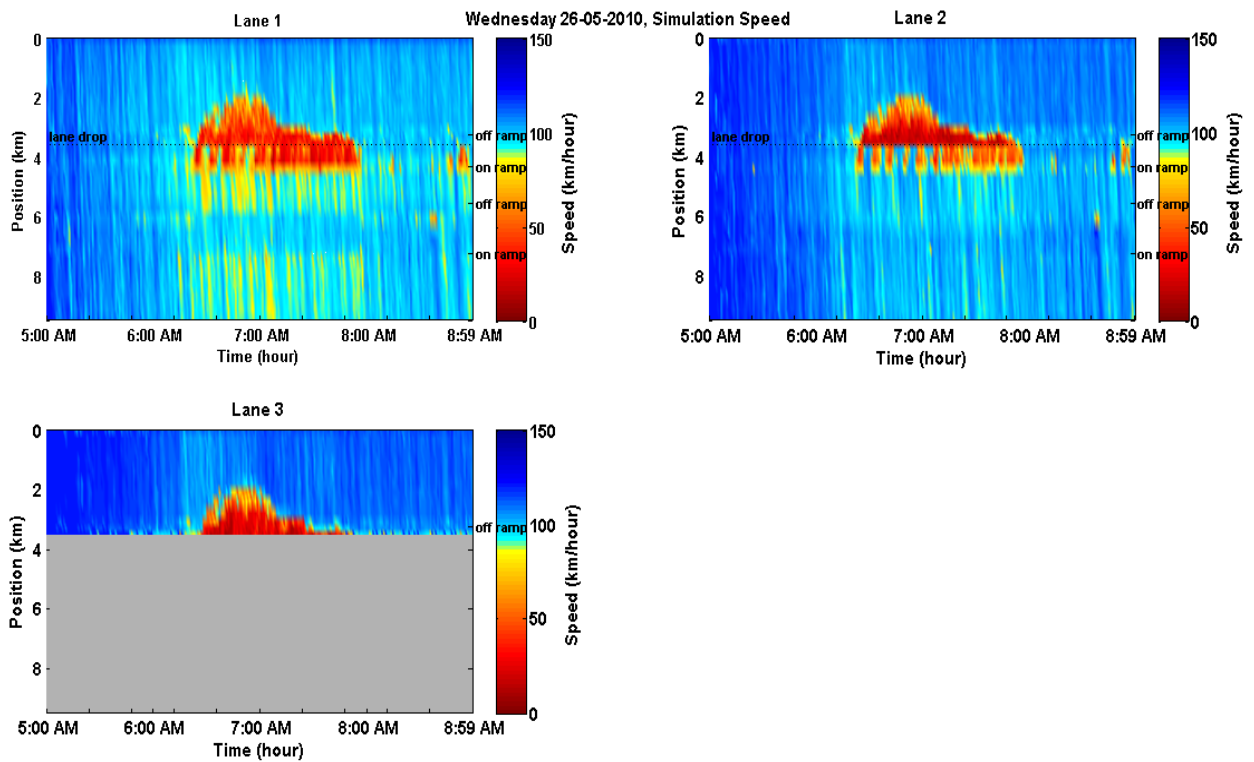

Figure 3.2: The space/time evolution of calibrated speeds.The presence of the lane drop after the first off-ramp marked as a dotted line is visible [10].

As we can witness from the above plots, congestion for all lanes of the network starts approximately at 6:50 AM (because of increased flow of vehicles entering the On- Ramp "Nieuwerkerk a/d IJssel" and finishes at 8:10 AM.

## 3.4 Realization of MOBIL strategy

After we have set the necessary background of our simulation and our calibrated network, we are in position to venture further to the impact of MOBIL within our specified network. In order to attain a more complete description of MOBIL behaviour to our calibrated network, I ran certain simulations for different combinations of politeness factors and thresholds, as depicted in (2.2). Specifically I ran some simulations for politeness factors (0.4, 0.8 and 1.0) and thresholds (1.0, 2.0 and 2.5). Furthermore, I tested the behaviour displayed for different regions of the network. This implies the free-flow area, lane-drop area and the congested area. Here in the following figure we can see ,how the MOBIL works based on the measurements received at each time step for both directions (left and right).

At first MOBIL computes the aggregated accelerations computed from the Incentive Criterion in Equation (2.2) for both prospective directions ( left and right), should the vehicle be in the median lane of our network (2). Our network is composed by three lanes. Otherwise if the vehicle is in the leftmost or the rightmost lane of the network, then only one prospective direction is taken into account. If one direction gives greater prospective accelerations than the other and should the Gap be sufficient to avoid a collision the the MOBIL controller performs a lane change at the next time step. The same happens the other way around for the other direction. And the measurements are being updated every 0.4 seconds, which is the time step of our simulated environment in Aimsun.

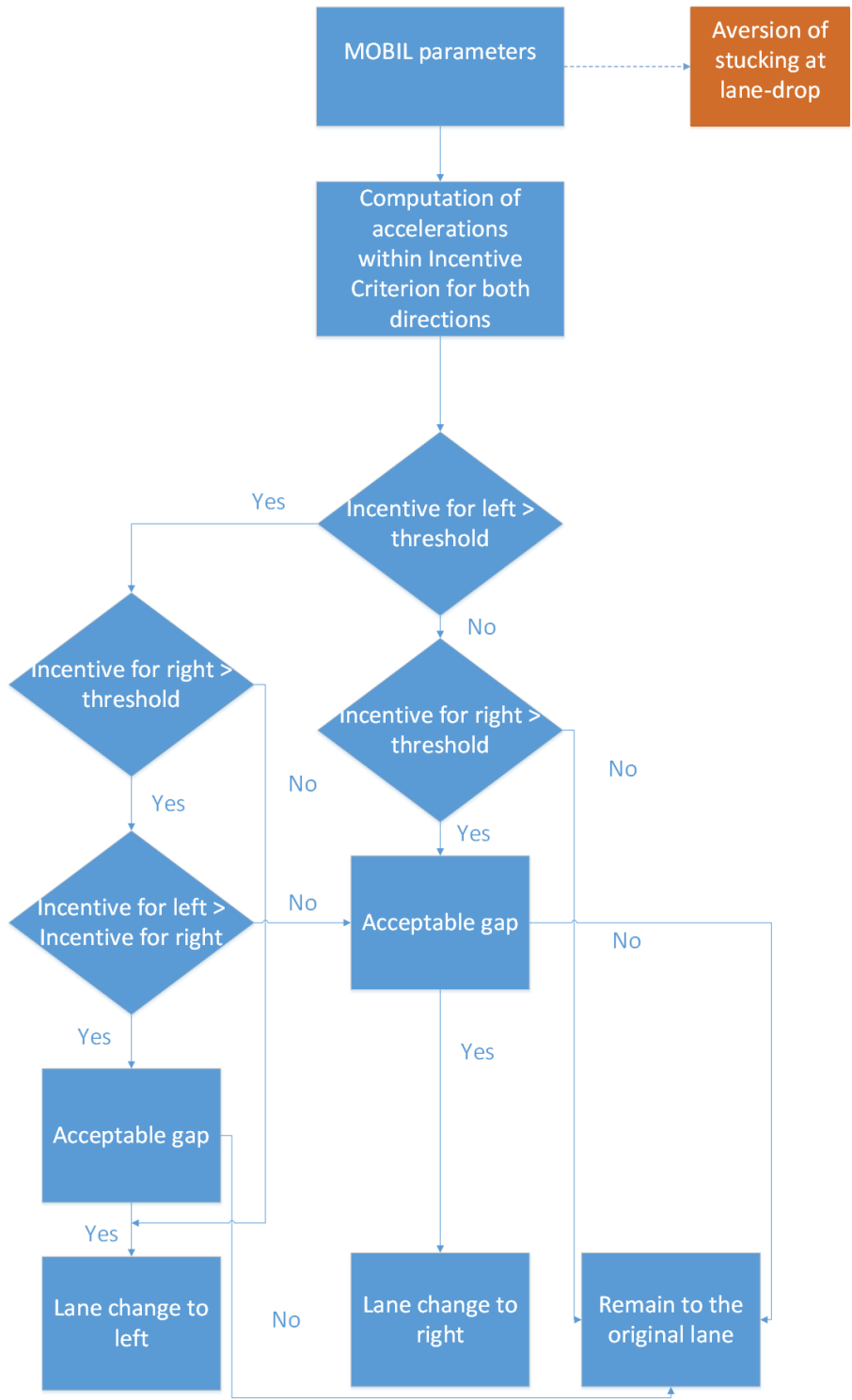

Figure 3.3: Graphical representation of MOBIL strategy

### 3.4.1 Impact of various thresholds and politeness factors

Here we examine the behaviour of MOBIL under different circumstances. First of all, we are going to test the free-flow conditions.

#### 3.4.1.1 Free-flow conditions

We inject a vehicle with MOBIL in the beginning of the free-flow area. As free-flow, we consider the time spectrum between 5:00AM - 6:49AM. Let us discuss the impact of politeness factor 0.4 and threshold 1.0 in Figure 3.4 .

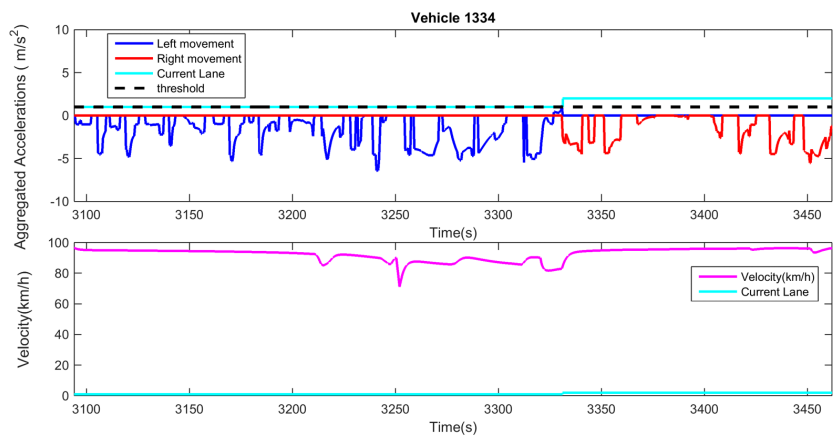

Figure 3.4: Behaviour of MOBIL in the freeflow area under politeness factor 0.4 and threshold 1.0.

As we can see from the figure above, the x-axis represents the time, the vehicle (injected with MOBIL) enters the simulation until the moment it exits the simulation. The y-axis in the first plot illustrates the accelerations computed at each time step (0.4s) from the left hand side of (2.2).As far as the second plot is concerned, in the y-axis we observe the velocity of our vehicle at each time step as well. The moment our vehicle of interest, enters the network its velocity is around 100 km/h. As time progresses, the vehicle's velocity decreases, indication that our car meets more frequently at each time step slower cars in front of it. From this perspective, our vehicle is obliged to break more often in order to avert any accidents with it's preceding vehicles.

What we assess in MOBIL, is the tendency of the vehicle to change lane to the right direction as long as the tendency of the vehicle to change lane to the left direction. Should a vehicle be in a median lane (in our particular network: lane 2), we compute both the tendencies, through the incentive criterion mentioned in Chapter 2.3.1. Then, the vehicle injected with MOBIL , changes lane towards the one , which has the grater incentive (computed from the sum of all the accelerations at (2.2). Should a vehicle be on the rightmost lane (lane 3), or at the leftmost lane (lane 2), then only one incentive is computed at each time.

The black dashed line represents the threshold imposed (1.0) for this case, the blue line represents the incentive of the vehicle to move to the left lane, the red line refers to the incentive of the vehicle to move to the right lane and the cyan line is admitted to the current lane in which the vehicle ( affected by MOBIL), is situated at each time step.

Each time, if one or both lines (red or blue) , exceed the threshold, then the vehicle changes lane at the next time step towards the lane , which has greater incentive. The reason of lane changing, is to improve firstly the traffic condition of our vehicle (gain in speed after the lane change) and secondly the traffic conditions of the neighbouring vehicles.

We can see, that approximately at 3325s, after a lane change towards the left direction our vehicle gains in speed , improving it's current traffic state and maintaining a increasing speed until it exits the network.

However, as we will witness from other cases as well, there is a factor that is hindered, which could affect the lane changing process. This factor is called *required gap* and refers to the distance of our current vehicle compared with the putative leader and follower in the target lane. Even though the threshold (to change lane) is exceeded, based on (2.2), there is a possibility, that our car collides with another vehicle in the target lane in a prospective lane change. So as to avoid this kind of behaviour, we have imposed gaps, so as to ensure that vehicles do not conflict with each other. This notion would be better comprehended with the following Figure 3.5 taken at some point from this network:

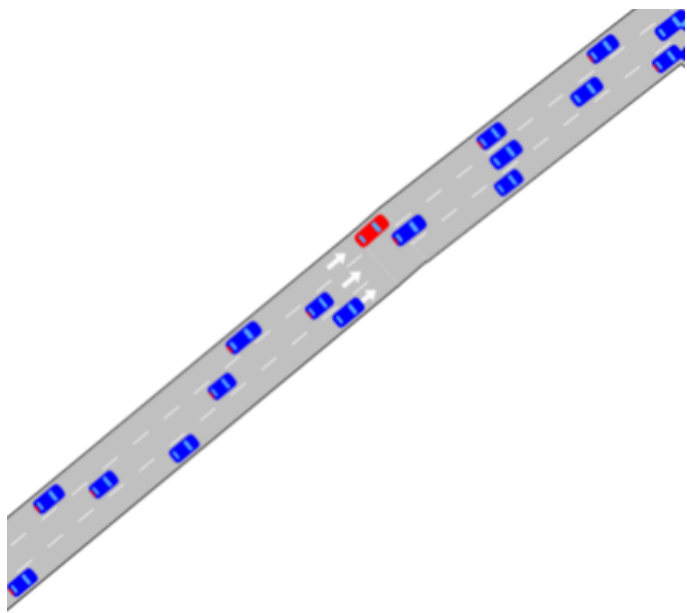

Figure 3.5: Conflicting goals for vehicle injected with MOBIL (vehicle with red colour)

This behaviour is manifested firstly for the left movement. What we can realize from Figure 3.4, is the value of, of the aggregated accelerations ranging from [-7,0], for a significant period of time (until 3345s). This result stems from the fact of avoiding conflicts after a lane change. So, at each time step, the vehicle computes the projected accelerations for both directions(left and right). The reason, that it doesn't exceed zero meausurements (until time 3345s) , is the aversion of collisions.

Let us now examine some other combinations of politeness factor and threshold for this specific region.

It is intriguing to check the behaviour of MOBIL model in the freeflow area for rather small values attached to politeness factor and threshold imposed, based on Equation (2.2). Figure 3.7 gives us a taste.

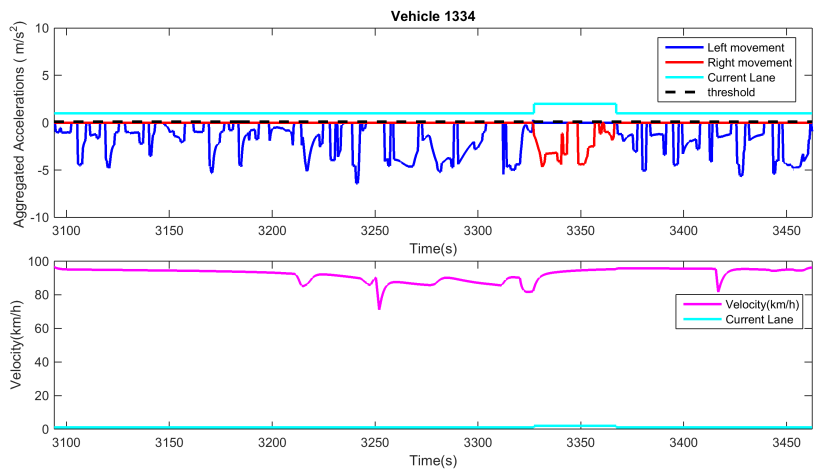

Figure 3.6: Behaviour of MOBIL in the freeflow area under politeness factor 0.4 and threshold 0.1.

As depicted in the above figure, there are just two lane changes taking place in the whole spectrum (that our vehicle is injected with MOBIL) of the network. This fact can be explained to the significantly reduced threshold imposed (from 1 to 0.1). This threshold is easier to exceed for vehicles, than this of 1.0, in the previous case, however the politeness factor (0.1) isn't enough to ensure that at each time step, we will have aggregated accelerations greater that 0.4.

At this point let's examine what happens, when we impose different parameters for the incentive criterion in equation (2.2).

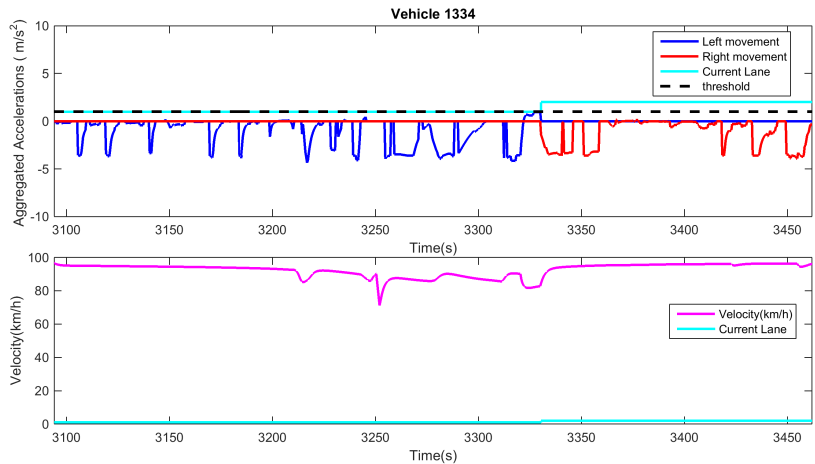

Figure 3.7: Behaviour of MOBIL in the freeflow area under politeness factor 0.1 and threshold 1.0.

The behaviour of this vehicle attached to MOBIL logic, is very much alike to the one manifested in Figure 3.7 The diminished politeness factor leads to diminished values of aggregated accelerations. However, these values prove to be just enough to surpass the threshold as displayed at time 3330 seconds.

### 3.4.1.2 Lane-drop conditions

We now check the behaviour of another vehicle (compared to the free-flow area). This vehicle has ID 3271. We incorporate this case as well, because the vehicle at some point gets stuck at the lane-drop ( time 4980-5020s). It is interesting to provide a thorough explanation of this tendency of our vehicle.

Each vehicle demonstrates a different behaviour, when injected with MOBIL. The reason is the properties of the vicinity vehicles ,not only in the current lane but also in the target lane as well, which could affect the performance of the vehicle we are examining (MOBIL). In this specific case, the computation of the accelerations based on (2.2) at each time step (for the vehicle attached with the MOBIL logic), taking into account the accelerations of the putative leader and follower in the target lane determine, how the vehicle will behave. Our vehicle is forced at some point, based on (2.2), to remain for a fraction of time at the lane-drop, rendering the traffic conditions more convenient for the rest of the vehicles of the network (not affected by MOBIL). MOBIL formulation is implemented not only to ameliorate the traffic state of the vehicles attached with its logic, but also to accommodate for the vehicles not conformed with its logic. By doing so, we improve the traffic state conditions of the whole network, leading steadily to a more solid and balanced environment.

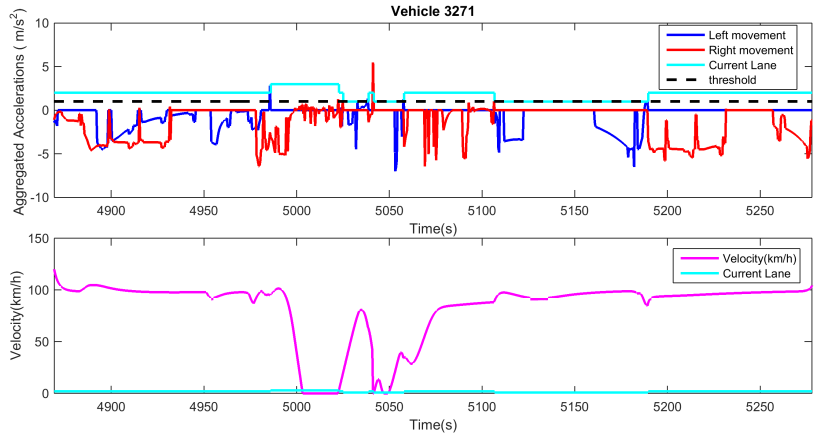

Figure 3.8: Behaviour of MOBIL in the lane-drop area under politeness factor 0.4 and threshold 1.0.

As we can observe from 3.8, vehicle 3271 enters the network at time 4850s and exits the network at time 5250s. So, its existence at the lane drop is of trivial significance, since it doesn't affect vehicle's 3271 performance to a large scale. The moment the vehicle 3271 ascends to the lane drop area, its velocity is plummeted rapidly, until it reaches zero, where the vehicles is situated right on top of the lane-drop (sole leader), with no movement whatsoever.

The vehicle for a period of approximately 40 seconds is unable to find proper circumstances, based on the vicinity vehicles and the available gap to perform a lane change. However, after 41 seconds, the vehicle finds a sufficient gap and is relieved from the lane-drop area. It is reasonable, that after it exits lane drop, our vehicle injected with MOBIL (3271) gains in speed until if finds a platoon of stopped vehicles in front of it and in the adjacent lanes. This can be illustrated at point 5025s, where the vehicle's speed starts reducing again, reaching nearly zero.

As in the previous case, after a lane change occurs, the vehicle gains in speed signifying, that the vehicle behaves in such a way to favour it's traffic state in the target lane. As the vehicle reaches the exit of the network it picks up on speed, due to the smaller amount of vehicles situated at that point.

Furthermore, we are going to incorporate the behaviour of vehicle 3271, with a combination of low values (politeness factor and threshold). Here are our experiment results provided by Aimsun.

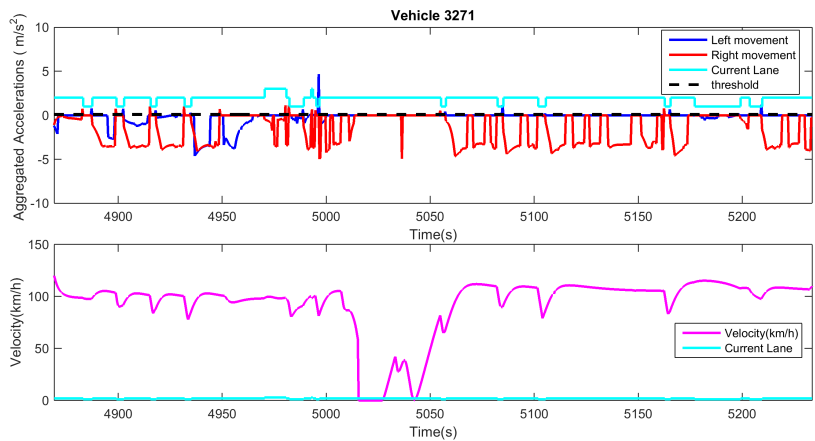

Figure 3.9: Behaviour of MOBIL in the lane-drop area under politeness factor 0.1 and threshold 0.1.

As we can see from Figure 3.9, there is a significant increase in the number of lane changes, compared to the ones that are being displayed in Figure 3.8. What must be mentioned here, is that with such low values of politeness factor and threshold, our vehicle doesn't get stuck in a lane-drop region. It may be for the time frame of (5000-5060) seconds to lane 2, but watching the velocity curve, our vehicle does have some points, that the velocity isn't zero. There are some fluctuations in the velocity, during the vehicle's 3271 route. At time 5030 seconds, the vehicle gains in speed, since it doesn't encounter any vehicle in front of it , but later on it reduces speed so at to avoid accidents with vehicles, that are in a close range from 3271. The velocity seems to be increasing after a lane change ,until our vehicle finds itself in a close range with the vicinity vehicles (where velocity drops again).

### 3.4.1.3 Congested conditions

In this case we will examine the behaviour of a vehicle situated in the congestion period of the network (6:30 - 7:10 AM). The vehicle we will subdue to simulations is the one with ID 5265. In Figure 3.10, we can witness, how the vehicle responds to the injection of MOBIL.

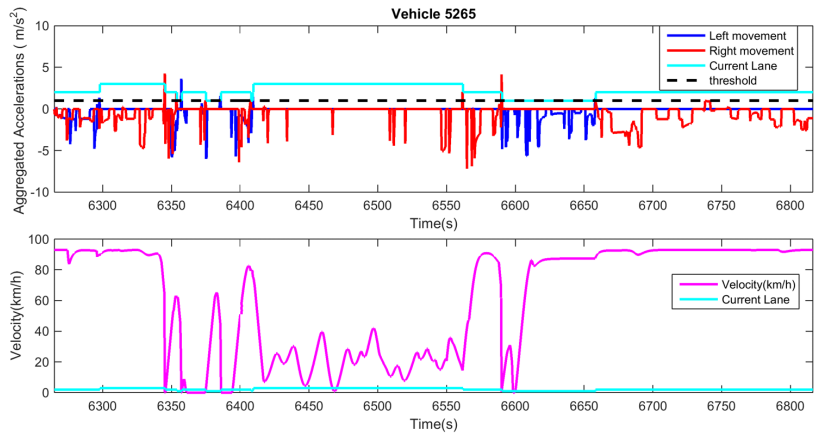

Figure 3.10: Behaviour of MOBIL in the congested region under politeness factor 0.4 and threshold 1.0.

The time our vehicle was affected by MOBIL was nearly 600s. In this region of our network, one element that needs to be taken into account is the more frequent number of lane changing that are being performed. Moreover, there is a significant increase in oscillations of speed , compared with the other two regions mentioned earlier. The velocity initiates from a value of nearly 95km/h, shortly after the vehicle enters the simulation network. Due to congestion, at time 6350s, there lies a very sudden drop in the velocity reaching zero. During it's route, our vehicle confronts many *stop and go waves*, which force it to immobilize and pick up on speed subsequently.This is illustrated as well in the time frame of (6410-6560) seconds, where although our vehicle lies within a specific lane with no lane change whatsoever, still there are many oscillations of speed. Regarding the range of velocity drop and immediate gain, in the *stop and go waves*, we can observe a rather smaller drop, than with the velocity right before and right after a lane change.

Here we do include another one figure , which shows the correlation between medium scale politeness factor and low threshold.

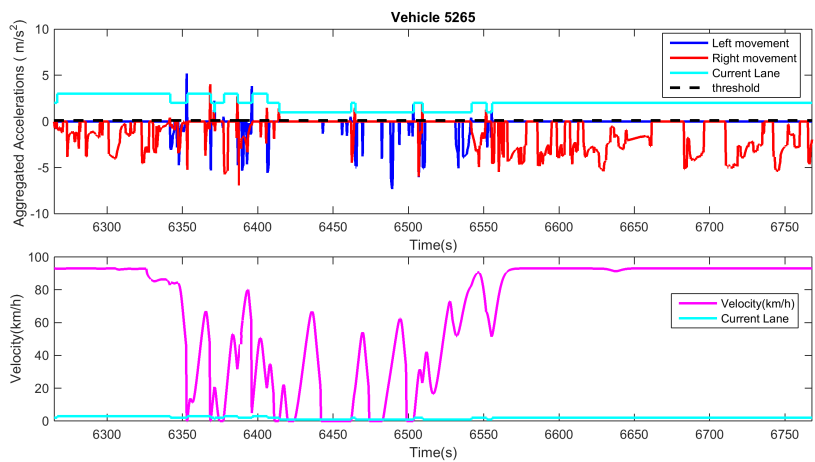

Figure 3.11: Behaviour of MOBIL in the congested area under politeness factor 0.4 and threshold 0.1.

Compared to the other two areas of the network (free-flow and lane-drop), in this area we have more incidents of lane changing occurring. This is reasonable, since our vehicle, which is injected with MOBIL, meets more cars in it's route and at each time step there is a higher probability of exceeding the threshold imposed, depending on the parameters set on Equation (2.2). This specific vehicle enters the network at time 6250 seconds and exits the network at 6800s. As we can see from the time range, that our vehicle is found within the network, it requires more time for our vehicle to be injected with MOBIL (due to the congestion taking place). As far as the velocity oscillations are concerned, we can see that at the time frame between [6350,6520], they are particularly tense and they are increasing rapidly and are being dropper rapidly accordingly. This is cause due to the congestion that propagates back in the network affecting our vehicle as well. In addition, *stop and go waves*, could be a source of this incident for instant velocity drop. In the same manner, the vehicle after being stopped, it gains in speed very quickly after being unstuck. This is a trait of MOBIL and if our vehicle finds the right circumstances and exceeds the threshold imposed on Equation (2.2) at a time step, then this could be an indicator that at some points it may avoid traffic and direct to the acquisition of a more favourable traffic state in the network.

We incorporate another one situation in order to have a more broader understanding of the behaviour manifested by the same vehicle as before, but with different parameters.

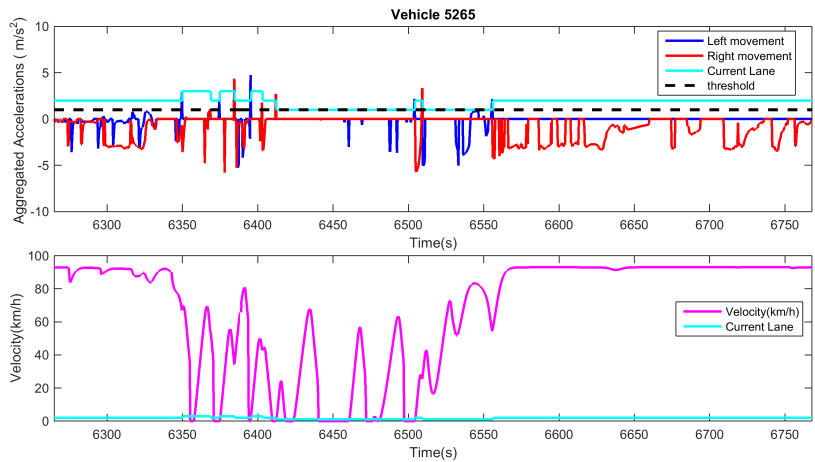

Figure 3.12: Behaviour of MOBIL in the congested area under politeness factor 0.1 and threshold 1.0.

The behaviour is similar with the one displayed at figure 3.10. A different trait of this set of politeness factor p and threshold $\Delta a_{th}$ is revealed in the above figure. During the vehicle's route around time (6350-6600) seconds, where multiple lane changes take place, the vehicles acquires greater values of velocity at times ( compared to the Figure 3.10) , which as in the Figure 3.10, drop due to the frequent *stop and go waves*.

Now we are going to extend our results for other combinations of politeness factor and threshold. Particularly, (see Figure 3.13).

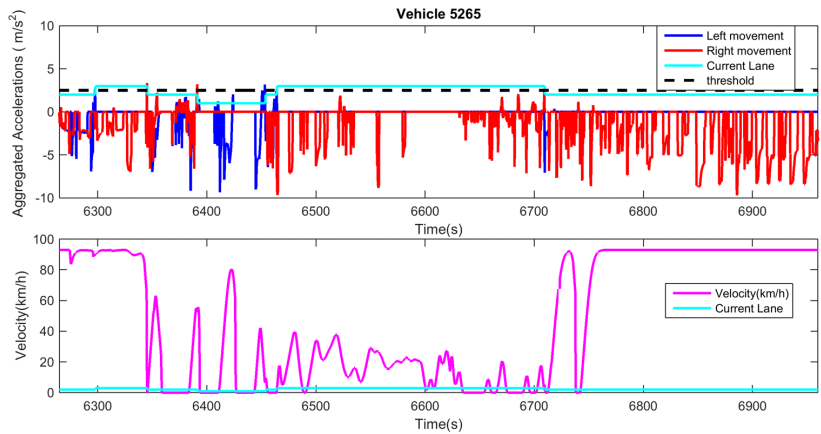

Figure 3.13: Behaviour of MOBIL in the congested area under politeness factor 0.8 and threshold 2.5.

As we can see from the Figure above, we have imposed a rather high threshold, which enables a not so comfortable and frequent lane change.

The first thing, is that the time spectrum, in which our vehicle is injected with MOBIL is 700 seconds (6250-6950) seconds. This means that the simulated vehicle needs additional 150 seconds compared to the similar case (under different circumstances) in Figure 3.10 to exit the simulation. This could be explained to the fact of a relative high threshold, which doesn't allow the vehicle to perform a lane change for a slow margin of accelerations. At each time step, our vehicles makes the computations based on the putative leaders and followers in the target lane (for both directions) and , if it finds the correct conditions, it changes lanes instantaneously. However, with such a high threshold, it is unlikely a vehicle finds the right conditions. Thus, it finds itself among other vehicles, which are at occasions stopped or close to velocity zero and do not allow this vehicle to make a transition to another lane. That's why the velocity keeps reducing at this period of time. Vehicle gains in speed after the stopped vehicles in front of it , gain positive velocity and stop again in front of them, when the vehicles stop again. Of course, as the vehicles continues it's route more and more vehicles coming from the On-Ramps, approach the network, so it is more challenging for our vehicle (injected with MOBIL) to maintain a non-zero velocity route.

When the vehicle changes lane at time 6715 seconds, it gains very quickly in speed reaching value 90km/h. Obviously it meets downstream the network vehicles, which are stopped and since it had gained in speed, our vehicle respectively plummets in speed with a very high ryhthm.

Also , we will examine another one combination of those two parameters. This case is depicted in Figure 3.14

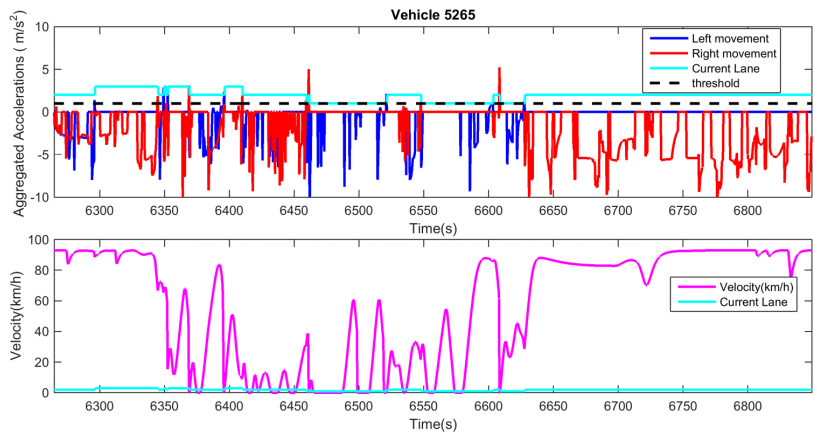

Figure 3.14: Behaviour of MOBIL in the congested area under politeness factor 1.0 and threshold 1.0.

The behaviour of the vehicle displayed in this Figure is much alike with the Figure 3.13. We can observe however, that due to the lower threshold imposed regarding the previous case ( threshold=2.5), there are more incidents of lane changing taking place. Also the oscillation of vehicles as far as the Figure 3.14 is concerned, we can detect getting greater values for the velocity , than those attained from figure 3.13. This happens , because the vehicle now doesn't confront such a big threshold to change lane and it is reasonable to change lanes more frequently, consequently avoiding long queues (congestion region).

# Chapter 4

# Evaluation of the travel time for different model parameters

We want to extend our analysis to evaluate the required travel time for vehicles injected with MOBIL strategy, to complete their route inside the determined network. We do not inject vehicles separately with MOBIL, so as to figure how its behaviour will be affected remotely, however we inject an array of vehicles. We ensure, that only after a vehicle within the array has finished its route, we inject the next one with MOBIL logic. This is necessary in order to guarantee stability and representativeness. The results we get for each array of vehicles, from a statistical point of view are sufficient, since each time we emulate the same replication scenario within the microscopic traffic simulator Aimsun. We choose vehicles arbitrarily and we exclude vehicles entering from an On-Ramp and vehicles exiting from an Off-Ramp. At the third array there seems to be an allocation of vehicles situated mainly in the congested region, while in the first two arrays, vehicles are mostly concentrated in the free-flow region.

We will need to assess the Travel Time taken for each of these vehicles. The interest of this work is to examine the behaviour of each of these vehicles mentioned later on for a variety of combinations, between politeness factor $p$ and threshold $\Delta \alpha_{th}$ mentioned in Incentive Criterion (2.2). More specifically we subdue each of these vehicles to a range of combinations among the following values for each parameter:

$$p \in [0.1, 0.2, 0.3, ..., 0.9]$$
$$\Delta \alpha_{th} \in [0.2, 0.4, 0.6, ..., 2.2]$$

## 4.1 First array of vehicles

We attach the array of vehicles that we used for the first run of the simulations in Table 4.1.

| Aimsun Results | | |
|---|---|---|
| Vehicle ID | Initial Lane | Simulation time vehicle enters |
| 846 | 1 | 5:38:46 |
| 1339 | 3 | 5:51:30 |
| 1755 | 2 | 6:00:10 |
| 2243 | 3 | 6:07:50 |
| 2678 | 1 | 6:14:24 |
| 3339 | 2 | 6:22:25 |
| 3976 | 3 | 6:29:10 |
| 4768 | 2 | 6:38:40 |
| 5449 | 1 | 6:47:55 |
| 6250 | 3 | 6:56:24 |
| 6852 | 2 | 7:06:10 |
| 7920 | 2 | 7:20:55 |

Table 4.1: Vehicles injected with MOBIL subsequently for the first array

We consider in general a 4h simulation scenario (5:00AM-9:00AM). Congestion commences at 6:30AM.

We include the results of several vehicles for all the possible combination of parameters depicted in a grid. We do not include all the grids of each vehicle, since some plots feature very similar behaviour. We will put two figures for each array with a varying behaviour. In the first place, we include the grid for vehicle with ID 846.

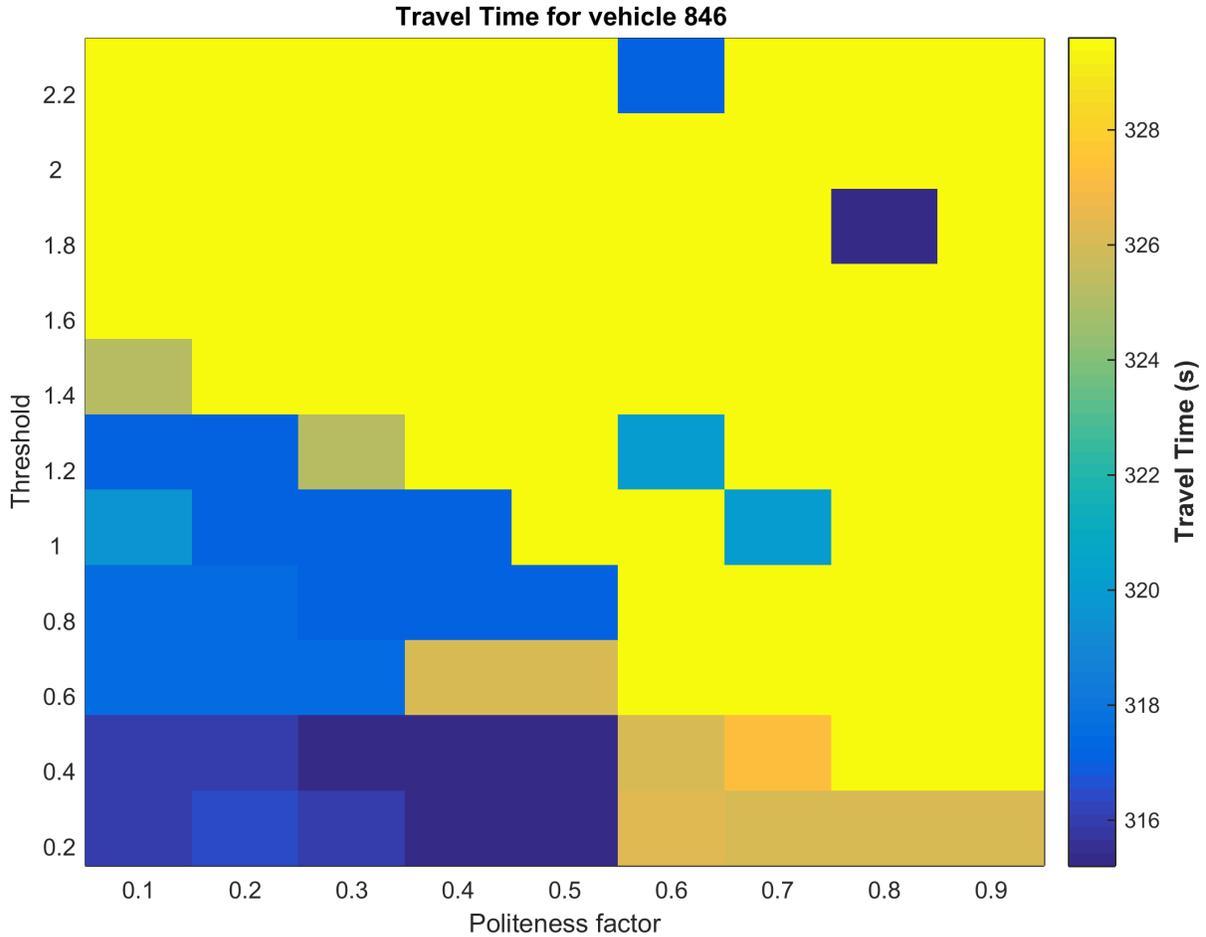

Figure 4.1: Travel time required for various combinations of two parameters for vehicle with ID 846.

As we can see from Figure 4.1, Travel Time seems to vary for different values of the two parameters imposed. These parameters are the ones mentioned in Incentive Criterion from Equation (2.2). For low politeness factor and for low threshold, the Travel Times maintains a low value. Well there are some remote cells in this grid which imply that there are some other combinations as well, which give low value of Travel Time as well. However, we want to provide a thorough explanation of the behaviour demonstrated by this vehicle in general. On the other hand we can see with yellow colour the area possessed by our vehicle, implying rather higher Travel Time within the network. Either for high threshold and high politeness factor or for low threshold and for high politeness factor the Travel Time maintains steadily the same value. Probably the vehicle doesn't respond any differently in this area rendering this case reliable.

Secondly we include the grid of vehicle with ID 2243.

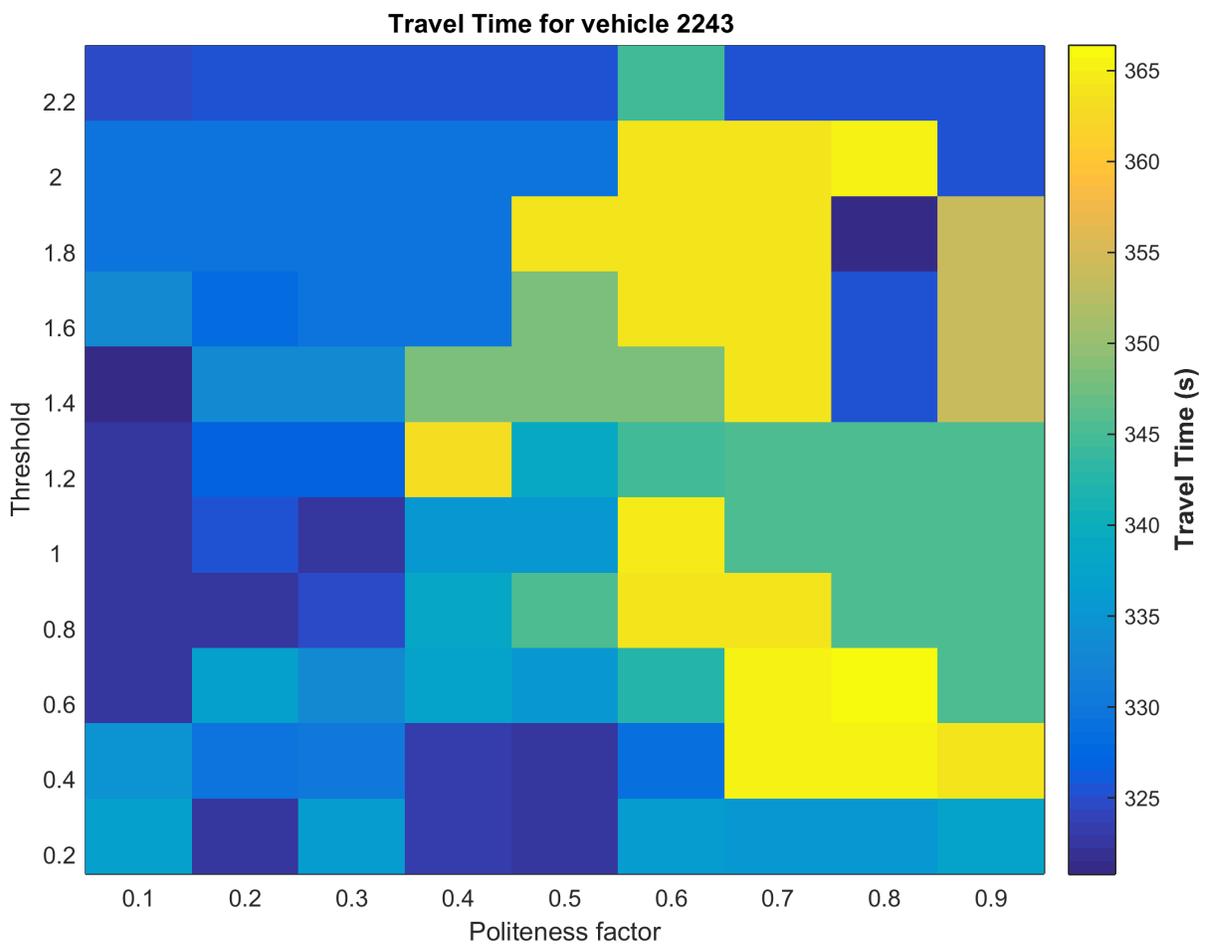

Figure 4.2: Travel time required for various combinations of two parameters for vehicle with ID 2243.

In Figure 4.1, we can detect that for high politeness factor and for high threshold, that at some occasions the Travel Time seems to have a higher value, that this of the other cells depicted. The opposite happens in the other region of the grid. For low politeness factor and for either low or high threshold , the distribution of Travel Time seems to be lower , than the other case. Our vehicle responds better to this "treatment".

## 4.2 Second array of vehicles

We continue the analysis within the scope of the second array of vehicles as described in Table 4.2.

| Aimsun Results | | |
|---|---|---|
| Vehicle ID | Initial Lane | Simulation time vehicle enters |
| 241 | 1 | 5:17:46 |
| 701 | 3 | 5:34:30 |
| 1011 | 2 | 5:44:27 |
| 1503 | 1 | 5:55:14 |
| 1978 | 2 | 6:04:14 |
| 2578 | 1 | 6:13:25 |
| 3482 | 3 | 6:23:10 |
| 4419 | 2 | 6:34:40 |
| 5133 | 1 | 6:42:55 |
| 5933 | 3 | 6:53:24 |
| 6894 | 1 | 7:06:10 |
| 7433 | 2 | 7:14:36 |

Table 4.2: Vehicles injected with MOBIL subsequently for the second array

We choose arbitrarily from the above array two cars. Firstly, we will investigate the behaviour of vehicle with ID 241.

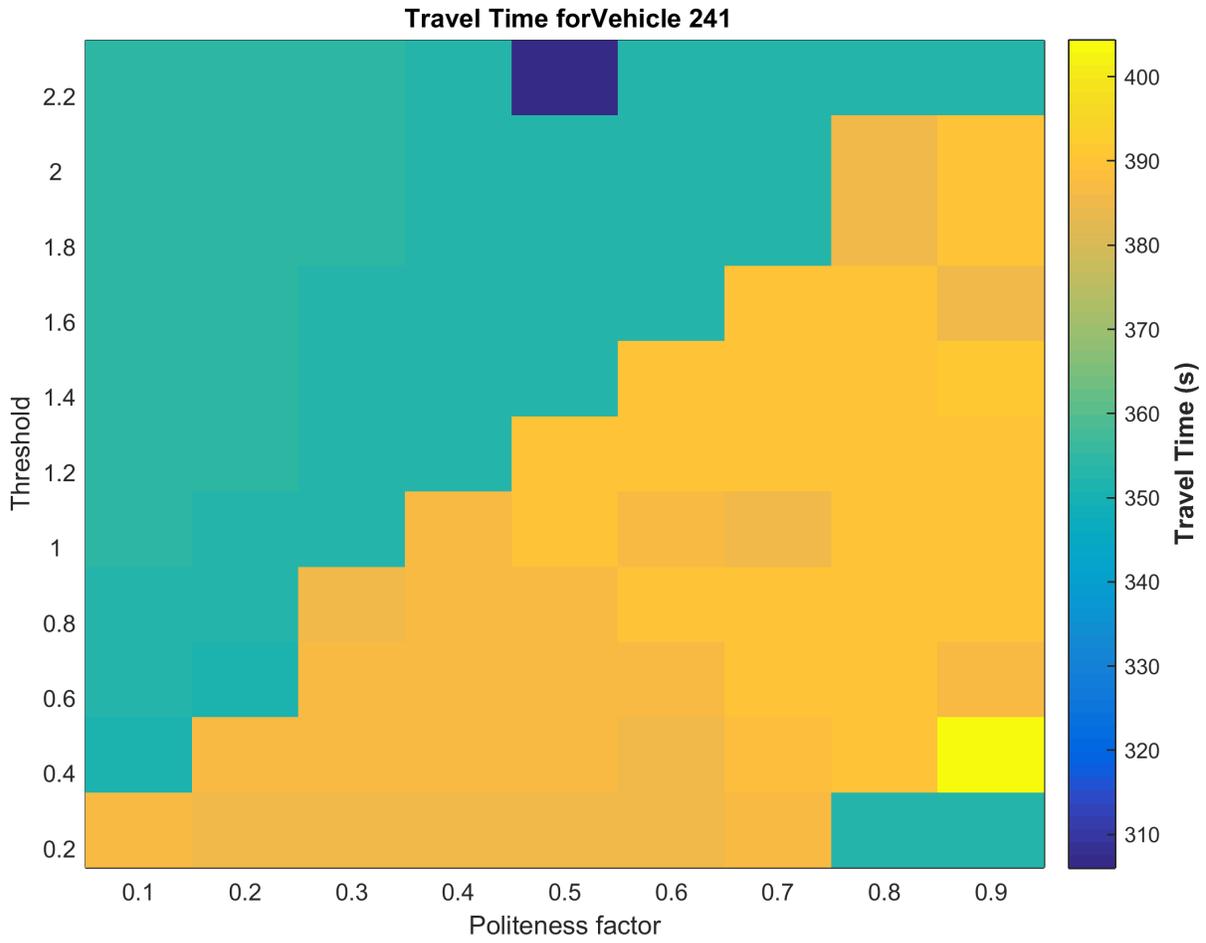

Figure 4.3: Travel time required for various combinations of two parameters for vehicle with ID 241.

What happens in the plot is quite obvious, as we can understand from the diagonal shape of the grid. We do not accomplish low Travel Times at all for this occasion. All cells are situated within 380-420 seconds.

So as to proceed we see what happens when we inject with MOBIL, vehicle with ID 5133.

Vehicle 5133 belongs in the congested region of the network. The distribution of times in this case is very promising, since only one value deviates from the general behaviour of our vehicle for this specific case. We can see, that we can accomplish rather low Travel Time for almost all the values, which ensures that our MOBIL controller works properly.

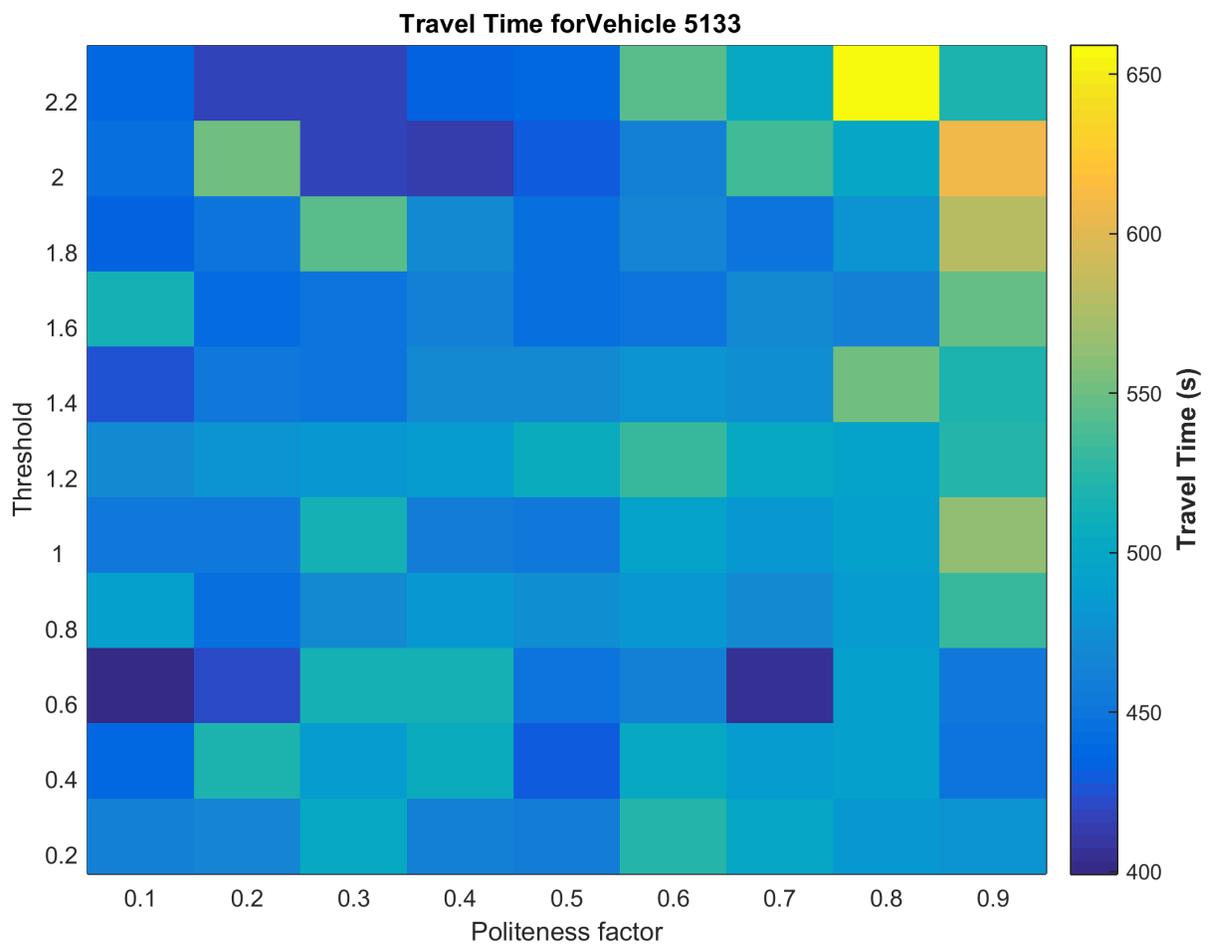

Figure 4.4: Travel time required for various combinations of two parameters for vehicle with ID 5133.

## 4.3 Third array of vehicles

We continue with the third array of vehicles in Table 4.3.

| Aimsun Results | | |
|---|---|---|
| Vehicle ID | Initial Lane | Simulation time vehicle enters |
| 2000 | 1 | 6:04:46 |
| 2653 | 2 | 6:14:40 |
| 3204 | 3 | 6:20:37 |
| 3821 | 2 | 6:27:14 |
| 4600 | 2 | 6:36:14 |
| 5243 | 1 | 6:44:25 |
| 5892 | 3 | 6:52:53 |
| 6493 | 2 | 7:00:45 |
| 7055 | 3 | 7:09:22 |
| 7679 | 1 | 7:17:55 |
| 8217 | 2 | 7:25:10 |
| 8724 | 3 | 7:31:40 |

Table 4.3: Vehicles injected with MOBIL subsequently for the third array

We choose to include the vehicle with ID 2000 for this particular case. Here is how it seems like.

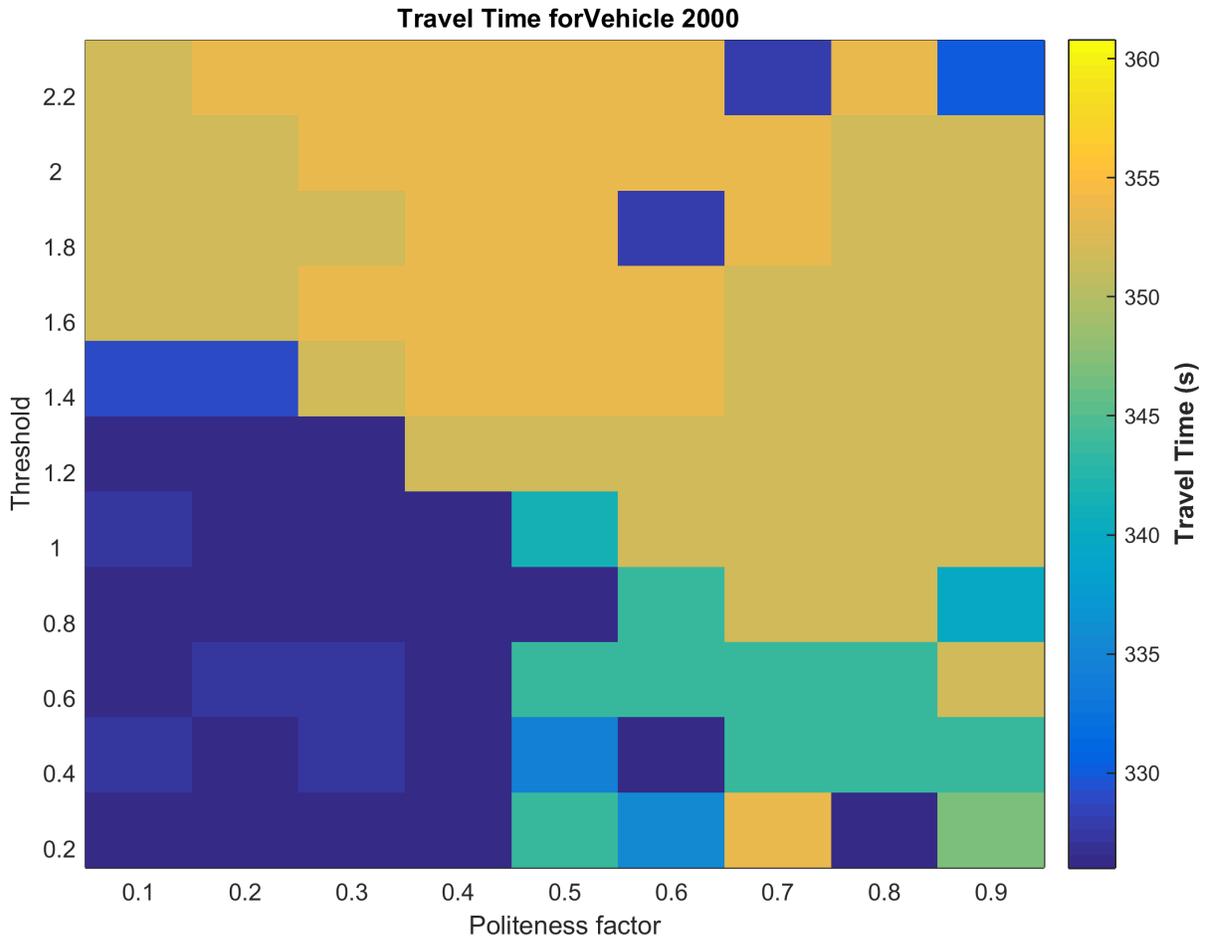

Figure 4.5: Travel time required for various combinations of two parameters for vehicle with ID 2000.

In that case there seems to be a pattern. This means that for low politeness factor and for threshold lower than 1.4, there is a distribution of times with low Travel Time. However, for high politeness factor and for threshold greater than 0.8, we do have bigger estimation of Travel Times. This is happening, because although we impose our vehicles with a high politeness factor, there is simultaneously a not so easy threshold to exceed. So our vehicle doesn't exceed the threshold many times, avoiding perhaps a more convenient traffic state.

We proceed with the injection of vehicle with ID 7055. Here is what we get:

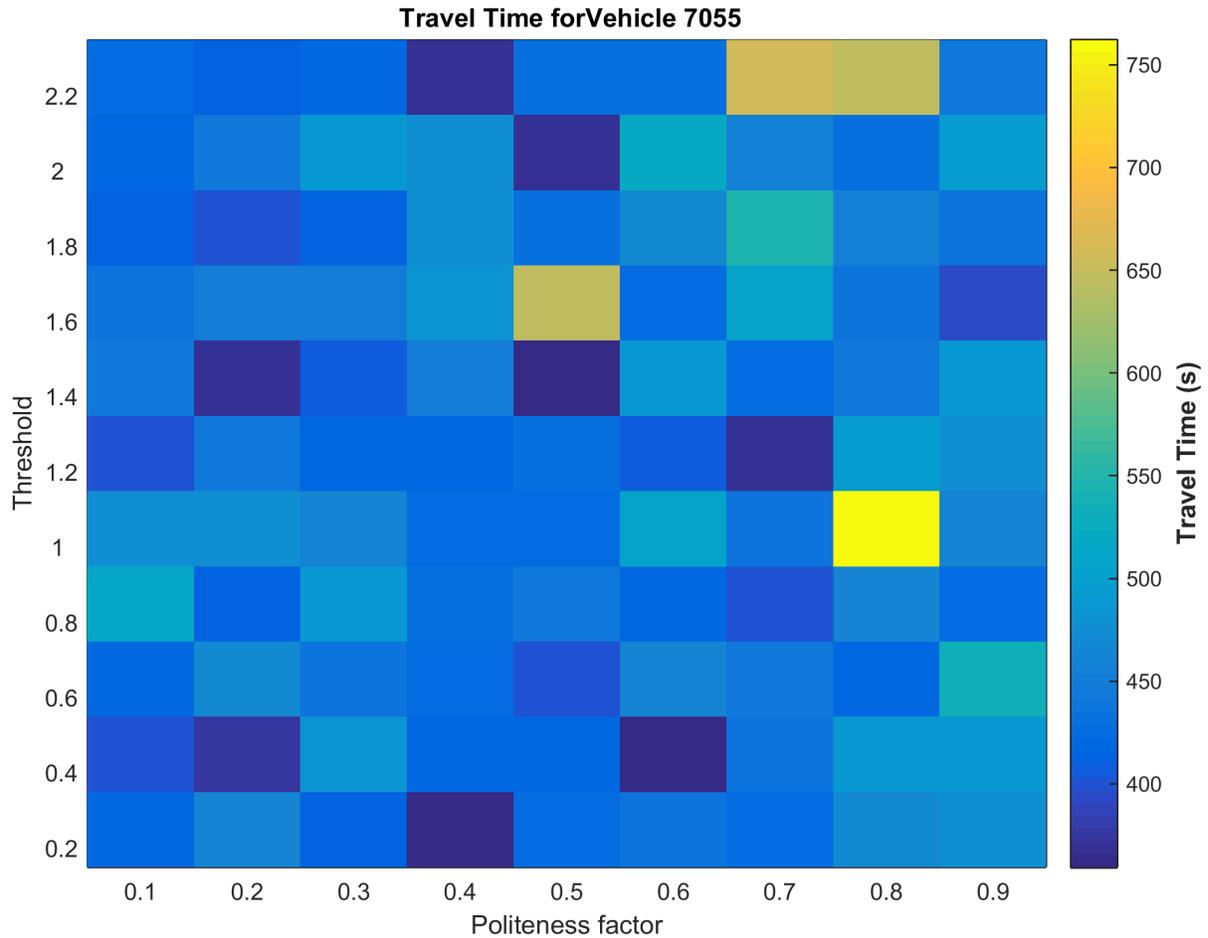

Figure 4.6: Travel time required for various combinations of two parameters for vehicle with ID 7055.

Again we can see a good distribution of Travel Times within our grid. It is important to mention, that our car is situated at this point in the congested area, so it's reasonable for the vehicle to take longer time to conclude its route, as we can see from the colourmap.

## 4.4 Fourth array of vehicles

We conclude our analysis with the fourth array of vehicles as mentioned in Table 4.4.

| Aimsun Results | | |
|---|---|---|
| Vehicle ID | Initial Lane | Simulation time vehicle enters |
| 1334 | 1 | 5:52:00 |
| 1952 | 3 | 6:03:50 |
| 2436 | 2 | 6:10:57 |
| 3424 | 3 | 6:22:25 |
| 4145 | 2 | 6:31:25 |
| 4970 | 1 | 6:40:40 |
| 6229 | 3 | 6:57:10 |
| 6998 | 2 | 7:08:40 |
| 8305 | 3 | 7:27:35 |
| 9126 | 1 | 7:36:35 |
| 10370 | 3 | 7:53:27 |
| 10993 | 2 | 8:02:15 |

Table 4.4: Vehicles injected with MOBIL subsequently for the fourth array

For this array, which is focused more on the congested region, we will pick one vehicle from the free-flow area and one vehicle from the congested area. What interests us is the difference in Travel Times which is contingent on the case we are examining at each time.

We start with vehicle with ID 3424:

What this plots illustrates is that the Travel Time can remain low even for high values of threshold and low values of politeness factors. This is very important since this could happen only based on the high aggregated accelerations estimated from Incentive Criterion, which are a function of the putative leader's and putative follower's accelerations.

In this plot , which is focused to the congested area, we can detect, that even though it is reasonable to allow greater Travel Times, there are quite a few combinations of parameters which give quite a low Travel Time in terms of the con-gestion. This is very beneficial for this car, because this proves that at times and under specific circumstances, our vehicle can perform the same way it would, in the free-flow , maybe even better.

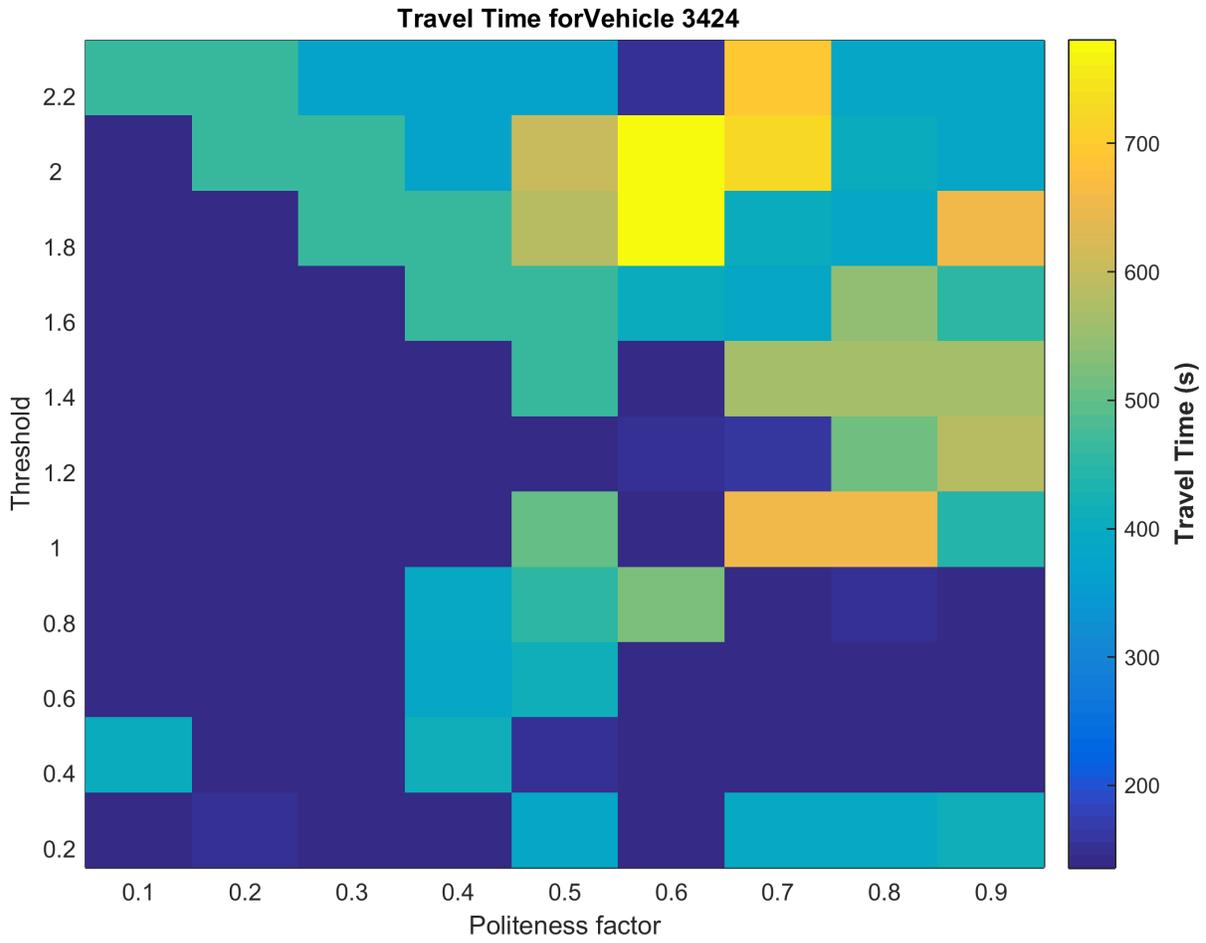

Figure 4.7: Travel time required for various combinations of two parameters for vehicle with ID 3424.

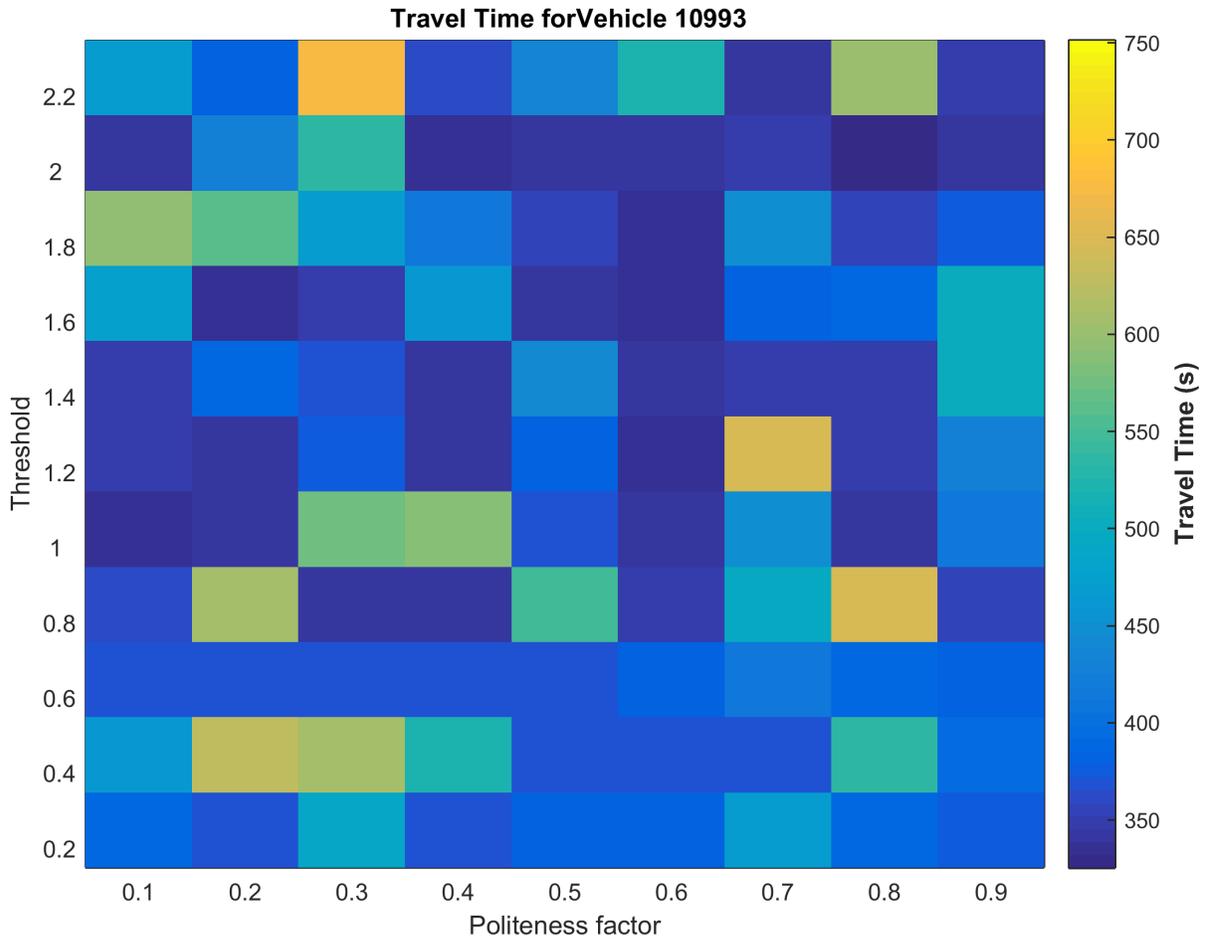

Figure 4.8: Travel time required for various combinations of two parameters for vehicle with ID 10993.

## 4.5 Average Travel Time for all vehicles

What we are going to do in this section is to estimate the Average Travel Time for vehicles belonging in the free-flow area of the network and the Average Travel Time for vehicles belonging in the congested area of the network.

Table 4.5: Allocation of vehicles

| Vehicle ID | Free-flow area | Congested area |
|---|---|---|
| 1334 | X | |
| 1952 | X | |
| 2436 | X | |
| 3424 | X | |
| 4145 | | X |
| 4970 | | X |
| 6229 | | X |
| 6998 | | X |
| 8305 | | X |
| 9126 | | X |
| 10370 | | X |
| 10993 | | X |
| 2000 | X | |
| 2653 | X | |
| 3204 | X | |
| 3821 | X | |
| 4600 | | X |
| 5243 | | X |
| 5892 | | X |
| 6493 | | X |
| 7055 | | X |
| 7679 | | X |
| 8217 | | X |
| | Continued on next page | |

| Vehicle ID | Free-flow area | Congested area |
|---|---|---|
| 8724 |   | X |
| 241 | X |   |
| 701 | X |   |
| 1011 | X |   |
| 1503 | X |   |
| 1978 | X |   |
| 2578 | X |   |
| 3482 | X |   |
| 4419 |   | X |
| 5133 |   | X |
| 5933 |   | X |
| 6894 |   | X |
| 7433 |   | X |
| 846 | X |   |
| 1339 | X |   |
| 1755 | X |   |
| 2243 | X |   |
| 2678 | X |   |
| 3339 | X |   |
| 3976 | X |   |
| 4768 |   | X |
| 5449 |   | X |
| 6250 |   | X |
| 6852 |   | X |
| 7920 |   | X |

Table 4.5 – continued from previous page

As we can see from the Table below we possess 22 vehicles in the free-flow area and 26 vehicles in the congested area. Some arrays focused more on the free-flow area , while others in the congested area. Those arrays are more or less in a balance and thus we expect them to give reliable results (like those mentioned in the previous arrays). So what we are going to implement now, is the grids both for free-flow region and for congested region in terms of the average Travel Time spent within the network for all the four arrays cumulatively.

We include the Average Travel Time grid for the free-flow conditions.

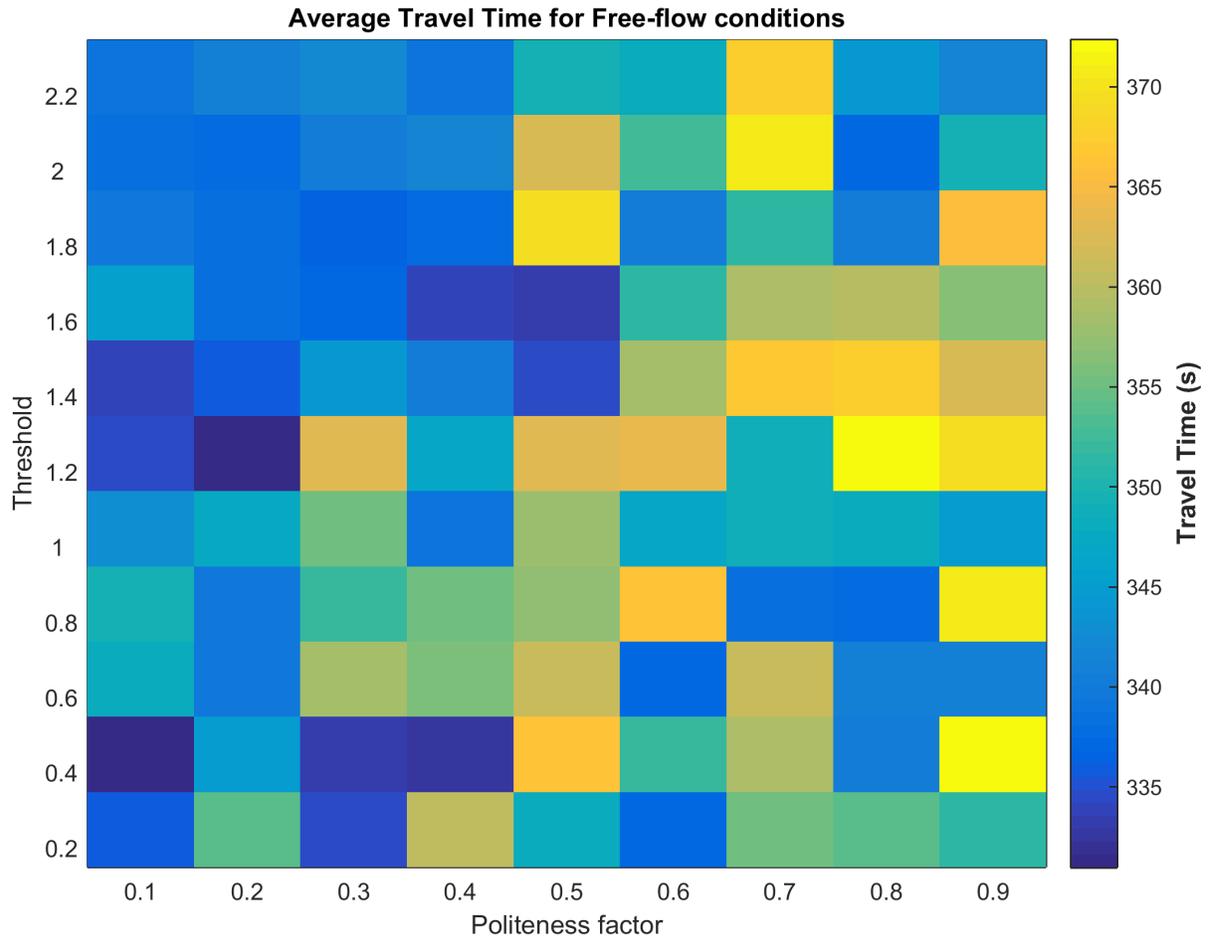

Figure 4.9: Average Travel time required for various combinations of two parameters for the free-flow area.

The results of the Average Travel Time for the free-flow are quite promising, since almost all the cells within the grid are charactesized by low Travel Times and thus very convenient traffic states for the vinicity vehicles of the ones adjusted with MOBIL controller.

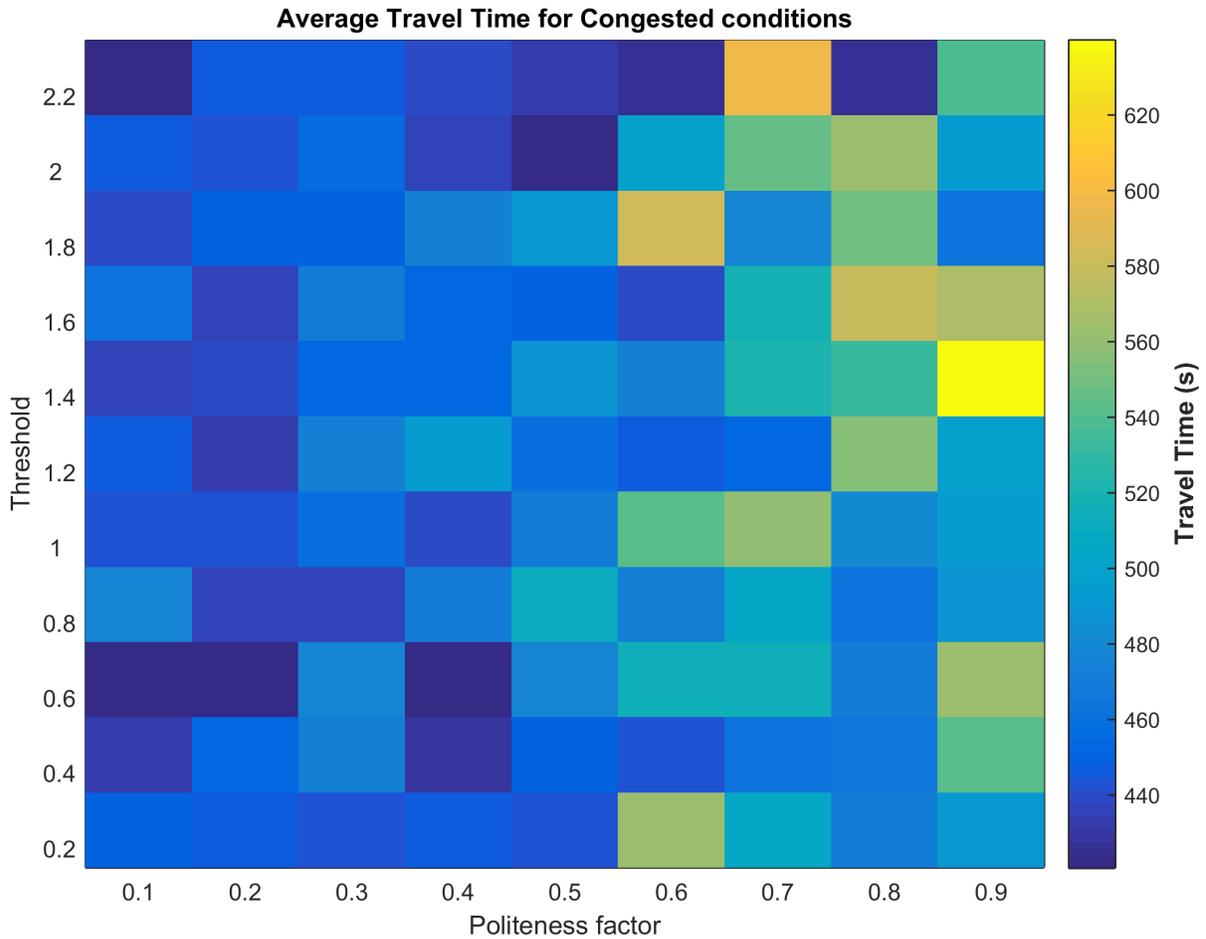

Figure 4.10: Average Travel time required for various combinations of two parameters for the congested area.

In this case as well, we can see a quite good behaviour as far as the vehicles, that represent the congested region. We can see that the range of Travel Time needed, for the vehicles to complete their route inside the simulated stretch in Netherlands doesn't take more than 650 seconds. Of course the vehicles in the congested area travel for a longer time , than those in the free-flow area, which is perfectly reasonable and signifies that the results we got are representative.

# Chapter 5

# Conclusions

Based on all the previously mentioned results we derived from the above experiments, we could deduce that, the MOBIL controller could be applied to vehicles enabling an automated or a semi-automated character rendering them more sophisticated. The results showed, that for every section of the network (free-flow, lane-drop or congestion area), MOBIL responds realistically based mainly on the incentive criterion, but also from the traffic state of the vicinity vehicles. Furthermore, as far as the Travel Time is concerned, this varies from vehicle to vehicle and is contingent on the traffic state of the specific vehicles we choose to inject with MOBIL. Anyway, the results above showed a good behaviour. What we must pinpoint, is that at some occasions, when high threshold was imposed, vehicles inevitably got stuck for a fraction of time in the lane-drop region, thus requiring further time to exit the network.